\documentclass[useAMS,usenatbib,12pt]{mn2e}
\usepackage[pdftex]{graphicx,color}
\usepackage[latin1]{inputenc}
\usepackage{graphics}
\usepackage{amsfonts}
\usepackage{amsmath}
\usepackage{multicol}
\usepackage{layout}
\usepackage{amssymb}
\usepackage[a4paper,colorlinks=true,pdfstartview=FitV,
linkcolor=green,citecolor=blue,urlcolor=magenta]{hyperref}
\title[The Substructure Hierarchy in Dark Matter Haloes] 
      {The Substructure Hierarchy in Dark Matter Haloes}   
\author[C. Giocoli et al.]
{\parbox{\textwidth}{Carlo Giocoli$^{1}$  
\thanks{E-mail: \href{mailto:cgiocoli@ita.uni-heidelberg.de}
	{cgiocoli@ita.uni-heidelberg.de}},
    	Giuseppe Tormen$^{2}$, Ravi K. Sheth$^{3}$,
    	Frank C. van den Bosch$^4$}   \\  \\
  	$^{1}$Zentrum f\"ur Astronomie, ITA, Universit\"at Heidelberg,
  		  Albert-Ueberle-Str. 2, 69120 Heidelberg, Germany    \\
  	$^{2}$Dipartimento  di  Astronomia,  Universit\`{a} degli  Studi  di
  		  Padova, Vicolo dell'osservatorio 2 I-35122 Padova, Italy \\
  	$^{3}$Center for Particle Cosmology, University of Pennsylvania, 209
  		  S. 33rd Street Philadelphia, PA 19104-6396\\
  	$^4$Department of Physics \& Astronomy, University of  Utah, 115
  		South 1400 East, Salt Lake City, UT 84112-0830}
\begin{document}
\date{}
\pagerange{\pageref{firstpage}--\pageref{lastpage}} \pubyear{2009}
\maketitle
\label{firstpage}
\begin{abstract}
  We present  a new algorithm for identifying  the substructure within
  simulated dark  matter haloes.  The  method is an extension  of that
  proposed by \citet{tmy04}  and \citet{giocoli08}, which identifies a
  subhalo  as a  group of  self-bound  particles that  prior to  being
  accreted by the  {\it main} progenitor of the  host halo belonged to
  one and the same  progenitor halo (hereafter `satellite').  However,
  this definition does  not account for the fact  that these satellite
  haloes themselves may also  have substructure, which thus gives rise
  to sub-subhaloes,  etc.  Our new  algorithm identifies substructures
  at all levels of this hierarchy, and we use it to determine the mass
  function  of all  substructure (counting  sub-haloes, sub-subhaloes,
  etc.).  On average, haloes which formed more recently tend to have a
  larger  mass fraction in  substructure and  to be  less concentrated
  than average haloes of the  same mass.  We provide quantitative fits
  to these correlations.  Even  though our algorithm is very different
  from  that of  \citet{getal04}, we  too find  that the  subhalo mass
  function  per  unit  mass  at  redshift $z=0$  is  universal.   This
  universality extends  to any redshift  only if one accounts  for the
  fact  that host  haloes of  a given  mass are  less  concentrated at
  higher redshifts,  and concentration and  substructure abundance are
  anti-correlated.  This universality  allows a simple parametrization
  of the subhalo  mass function integrated over all  host halo masses,
  at any given time.  We  provide analytic fits to this function which
  should be useful  in halo model analyses which  equate galaxies with
  halo substructure when interpreting clustering in large sky surveys.
  Finally,  we  discuss systematic  differences  in  the subhalo  mass
  function that arise from different definitions of (host) halo mass.
\end{abstract}

\section{Introduction}

In  the  standard  scenario   of  structure  formation,  galaxies  are
surrounded by extended dark matter haloes, which form by gravitational
instability   seeded  by  some   initial  density   fluctuation  field
$\delta(\vec{x},z)$.  According to the simple spherical collapse model
\citep[e.g.][]{pee80}, a halo of mass  $M$ collapses at a redshift $z$
when the  linear density  fluctuation field -  smoothed on a  scale $R
\sim    M^{1/3}$   -   first    exceeds   some    critical   threshold
\citep{ps74,bond91,lc93,st02}.   Small   systems  collapse  at  higher
redshifts when the universe  was denser, and then merge hierarchically
to  form  larger haloes.   Galaxy  clusters are  at  the  top of  this
hierarchy: they  represent the  most massive virialized  structures in
the present-day universe, and  may host thousands of galaxies.  Recent
studies,     using    high     resolution     numerical    simulations
\citep[e.g.][]{m98,tds,swtk01}, have shown that the cores of subhaloes
accreted  along  the  host  merging  history  may  survive  until  the
present-time to form the so-called substructure population.

Different studies of galaxy  formation and evolution have attempted to
correlate such substructures with satellite galaxies, with conflicting
results.  $N$-Body  simulations of  galaxy-size haloes seem  to predict
more  substructures   than  observed  \citep{mooreetal99,setal03}.   A
number of different solutions to this `substructure problem' have been
suggested, among  others early cosmic  reionization \citep{susa}, mass
loss and  gas stripping \citep{ketal04,mac09},  and a downturn  in the
power spectrum at small scales \citep{kam00}.

From  the observational point  of view,  recent kinematic  analyses of
Milky-Way satellites  \citep{simonmw,walk} have shown  that, if galaxy
formation in low-mass dark  matter haloes is strongly suppressed after
reionization,  the   circular  velocity  function   of  simulated  CDM
subhaloes can be brought  into approximate agreement with the observed
circular  velocity  function  of  the  Milky  Way  satellite  galaxies
\citep[e.g.,][]{koposov09}.    \citet{Ishiy08,Ishiy09}  claim  instead
that  there is  no missing  satellites  problem if  one considers  the
scatter  in  the  substructure  mass  function due  to  the  different
formation history of the haloes.   The idea here is that the Milky-Way
and  Andromeda haloes  reside  within an  underdense  region, so  they
accreted their  substructures at above average  redshifts (compared to
other objects  of the same  mass today), and  so they are  expected to
possess fewer satellites \citep[see also][]{vtg05}.

Knowledge of  the substructure population  is important, not  only for
galaxy  formation  studies,  but  also  for  estimating  the  expected
observable $\gamma$-ray  flux from dark  matter particle annihilations
within  the Milky  Way.  The  expected flux  depends  strongly on  the
substructure population, its spatial distribution within the halo, and
on the DM particle cross section \citep{pbb,diemand07,gioc08}.

The observed thickness of stellar  disks in spiral galaxies is another
imprint    of    the    substructure    population   of    the    host
haloes. Semi-analytical models \citep{benson} and numerical simulations
\citep{kaz1,kaz2} show  that merging events  in the central  region of
the halo are responsible for disk thickening.  Hence, the substructure
mass function, its redshift evolution and the satellite accretion rate
all represent  key ingredients for a  more complete model  of the disk
structure of spiral  galaxies, not to mention flares,  bars, and faint
filamentary structures above the disk plane.

Different algorithms  have been proposed to  identify substructures in
simulated dark matter haloes.  All these algorithms yield subhalo mass
functions  that  resemble a  power-law  at the  low  mass  end with  a
logarithmic    slope   in    the   range    from   $-0.8$    to   $-1$
\citep{gill1,getal04,delucia04,shaw06,giocoli08,mad08,wetzel09},   both
at redshift $z=0$ and at  higher redshifts.  In this work we highlight
some differences  among four  such methods and  present a  new subhalo
finder which we believe is  useful for studies of galaxy formation and
evolution in that it is directly linked to the merger tree of the host
halo.

In this paper we focus on the study of the ``dark'' subhalo population
and on its redshift evolution, to help clarify the role of dark matter
and   its   dynamics  in   the   structure   formation  history.    In
Section~\ref{bkgnd}  we describe  several  substructure identification
algorithms already  in the  literature, including the  one we  will be
modifying.   Section~\ref{sim}  describes  the  cosmological  $N$-Body
simulation  we  use,  and   the  required  post-processing.   Our  new
algorithm  to  identify substructures,  and  an  analytic  fit to  the
resulting  subhalo  mass function  is  presented in  Section~\ref{r2}.
Section~\ref{r3} discusses  how the  scatter in the  substructure mass
fraction at fixed mass and  at different redshifts correlates with the
formation  history, and other  structural parameters  of the  halo.  A
summary and conclusions are given in Section~\ref{sc}.
In Appendix \ref{previous} we compare the results of this paper with 
two previous ones: \citet{giocoli08} and \cite{getal04} that studied the
substructures population on the same cosmological simulation but with 
different algorithms. In Appendix  \ref{massris} we propose
some fitting functions useful to rescale the virial radius and mass 
when different definitions of the enclosed virial density are adopted. 

\section{Background}\label{bkgnd}

To date,  several different algorithms have been  proposed to 
identify substructures  within  simulated  dark  matter  haloes,  
and  different definitions have been adopted by several authors.

\begin{itemize}

\item  \citet{ghigna2000}  used an  algorithm  called  SKID.  At  each
  simulation  snapshot, SKID  estimates the  density of  each particle
  using a cubic  spline kernel; each particle is  then moved along the
  density  gradients  until  it  oscillates  around  a  local  density
  maximum.  Particles are then linked using a friends-of-friends (FoF)
  algorithm  \citep{fof},  and  the   groups  thus  found  are  pruned
  iteratively to retain only self-bound particles.  However, to obtain
  the full  substructure mass SKID requires a  user-defined choice for
  the  linking  length  to  be   used  by  the  FoF  algorithm.   Also
  \citet{weller05}  implemented  an  analogous algorithm  to  identify
  substructures   by  moving  particles   up  density   gradients  and
  identifying  groups  as  all  particles reaching  the  same  density
  maximum.  In addition, they also  implemented a method to identify a
  hierarchy  of subhaloes  within subhaloes.   \citet{shaw07} improved
  the algorithm of determining the energetically bound components of a
  subhalo taking into  account also all the forces,  both internal and
  external, exerted on the subhalo.

\item \citet{swtk01}  developed SUBFIND, which  defines 'subhaloes' as
  locally  overdense,  self-bound particle  groups.   SUBFIND runs  on
  individual simulation snapshots,  but can afterwards reconstruct the
  full merger tree of each  subclump, by using the subhalo information
  from  previous snapshots \citep{speal05,croton,Hel06}.   The density
  of each  particle (assumed as tracers of  the three-dimensional dark
  matter density field) is  estimated using an SPH-fashion scheme: the
  local  smoothing scale  is set  to  the distance  of the  $N_{dens}$
  nearest neighbor, and the  density estimated by kernel interpolation
  over these  neighbors. Locally  overdense regions are  identified by
  imitating  such  a lowering  of  a  global  density threshold.   Any
  locally  overdense region  enclosed  by an  isodensity contour  that
  traverses a saddle point  is considered as a substructure candidate.
  All subhalo  candidates are also examined and  unbound particles are
  removed and redistributed.  This  algorithm has been used and tested
  by   different    authors   \citet{getal04,delucia04,speal05}.    In
  particular SUBFIND  has been run on the  largest existing simulation
  of the Milky Way \citep{acqua01,acqua02}.

\item \citet{gill1}  and \citet{gill2} used  MLAPM \citep{knebe01}, an
  adaptive  mesh  refinement  code  for cosmological  simulations,  to
  locate substructures in host  haloes.  To identify haloes and subhaloes
  at  each simulation  output, they  first build  a list  of potential
  centers by storing the centroid of the densest grid point at the end
  of each grid tree's ``branch''.   Assuming that each density peak in
  the adaptive  grid of MLAPM corresponds  to the centre of  a halo or
  subhalo, they step outwards  in (logarithmically spaced) radial bins
  until  the mean  enclosed  overdensity reaches  the critical  virial
  value  $\Delta_{\rm  vir}(z)$  \citep{eetal96,bn98},  thus  defining
  $R_{\rm  vir}$.   However,  subhaloes  do  not extend  out  to  their
  original virial radius, but rather to some truncation radius $R_{\rm
    trunc}$,  where  an  upturn  in  the  radial  density  profile  is
  detected.   This  rise  is  encountered  because  substructures  are
  embedded in the background of the host halo.

\item \citet{tmy04} and  \citet{giocoli08} developed and optimized the
  code SURV, which identifies subhaloes within the virial radius of the
  final host  by following their  progenitors from the time  they were
  first accreted by the host main progenitor. Hence, SURV differs from
  the methods discussed above in  that it uses prior information based
  on the merger tree of a  host halo at redshift $z_0$ to identify its
  subhaloes. For each progenitor halo  $p$ that has been accreted onto
  the  main progenitor,  SURV identifies  a subhalo  as the  subset of
  particles that belonged  to $p$ at the moment  of accretion ($z_{\rm
  acc} >  z_0$) and  are still  part of a  self-bound entity  at $z_0$
  within  the  corresponding  tidal  radius.  For  each  subhalo  thus
  identified,  SURV  stores the  accretion  time,  $z_{\rm acc}$,  the
  original virial mass  (i.e., the mass at $z_{\rm  acc}$), as well as
  the  evolution of its  orbital parameters  after accretion.  In this
  paper we  extend SURV by following  all branches of  the merger tree
  (rather than just the main branch), in order to reconstruct the full
  hierarchy  of  substructure  down  to  the mass  resolution  of  the
  simulation    (i.e.,    we    aim   to    identify    sub-subhaloes,
  sub-sub-subhaloes, etc).

\end{itemize}

\section{The Simulation}
\label{sim}

We used the data from GIF2 \citep{getal04}, a cosmological 
$N$-Body simulation which is available at
\href{http://www.mpa-garching.mpg.de/Virgo}
{\texttt{http://www.mpa-garching.mpg.de/Virgo}}.  The simulation
followed the evolution of $400^{3}$ dark matter particles, each of
mass $1.73 \times 10^{9}\,h^{-1}M_{\odot}$, in a periodic cube of
comoving side $110h^{-1}$Mpc within which the background 
cosmology was $\mathrm{\Lambda}$CDM with 
($\Omega_m$, $\sigma_8$, $h$, $\Omega_b h^2$) = (0.3, 0.9, 0.7, 0.0196).  
We made use of $50$ snapshots, mostly logarithmically spaced in time 
between $z=10$ and $z=0$.  These are sufficiently closely spaced in time 
that one can reconstruct accurate halo and subhalo merger trees 
\citep{tmy04,giocoli08}.  See \citet{getal04} and \citet{giocoli08} for more 
details about the GIF2 simulation and the post-processing.

\begin{figure}
\centering
\includegraphics[width=\hsize]{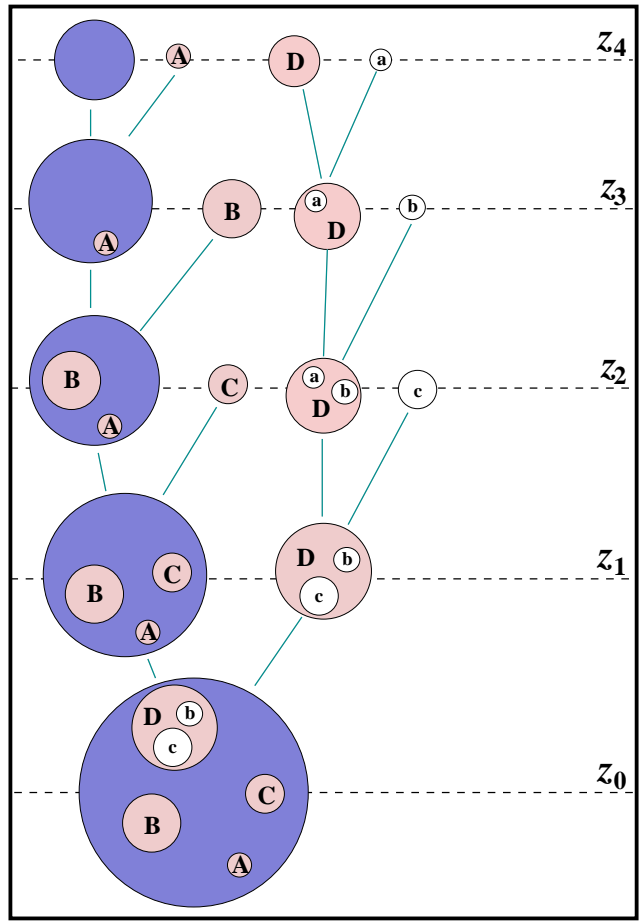}
\caption{Schematic representation of the merger tree of a dark matter
  halo along discrete time steps.  The dark-grey haloes on the left
  represents the evolution of the main halo progenitor, which accretes
  `satellite' haloes $A$, $B$, $C$ and $D$ that give rise to a
  population of subhaloes$^{(1)}$. System $D$ has accreted satellite
  haloes itself ($a$, $b$ and $c$), before it was accreted by the
  host. Those that survive ($b$ and $c$) give rise to a population of
  subhaloes$^{(2)}$.\label{treefig}}
\end{figure}

\subsection{Merger Trees and Substructures}

For each  simulation snapshot, haloes are identified  using a spherical
overdensity  criterion:  first  we  determine the  local  dark  matter
density  at the  position of  each particle,  $i$, by  calculating the
distance to the tenth closest neighbor, $d_{i,10}$. We assign to each
particle a local density $\rho_{i,\mathrm{DM}} \propto d_{i,10}^{-3}$,
sort particles  in density and  take as centre  of the first  halo the
position of the  densest particle.  We then grow  a sphere around this
centre, and stop  when the mean density within  the sphere falls below
$\Delta_{\rm vir}(z) \rho_{\rm crit}(z)$  as dictated by the spherical
collapse model.  For  a flat universe, this is  approximated to better
than one percent by
\begin{equation}
\Delta_{\rm vir}(z) = 9 \pi^2\, \left( 1 + \Omega_m(z)^{\beta} 
                                     - \alpha\,[1-\Omega_m(z)] \right) 
\label{felix}
\end{equation}
with  $\alpha=0.7076$ and $\beta=0.4403$  \citep{stth}.  For  the GIF2
cosmology,  $\Delta_{\rm vir}  =  97$ times  the  critical density  at
$z=0$,  and it increases  with redshift,  asymptoting to  $18\pi^2$ at
early  times. In Appendix~\ref{massris}  we present  fitting functions
that  allow halo  radius, mass  and concentration  to be  converted to
those appropriate for other values of $\Delta_{\rm vir}$.

At this point  we assign all particles within the  sphere to the newly
formed halo, and remove them from  the global list. We take the centre
of the  next halo at  the position of  the densest particle  among the
remaining ones, and grow a  second sphere.  We continue in this manner
until all  particles are screened.   We include in our  catalogue only
haloes  with  at  least  $10$  particles  within  the  virial  radius;
particles  not ending  up in  any halo  are considered  as  `field' or
`dust' particles.   Hereafter we  will refer to  the virial mass  of a
(host) halo thus identified at redshift $z$ as $M_z$.

\begin{table*}
\centering
\begin{tabular}{|l || c || c || c || c || c || c || c |} \hline
  observation redshift $z_0$& $11.5$-$12$ & $12$-$12.5$ & $12.5$-$13$ & 
  $13$-$13.5$ & $13.5$-$14$ & $14$-$14.5$ & $> 14.5$  \\ \hline \hline 

  $0$ & $8305$ ($\mathit{10230}$) & $3349$ ($\mathit{3897}$) & $1186$   
  ($\mathit{1346}$) & $461$ ($\mathit{503}$) & $127$ ($\mathit{141}$) & 
  $35$ ($\mathit{36}$)& $4$ ($\mathit{4}$)  \\ \hline 

  $0.5$ & $9347$ $(\mathit{11252})$ & $3544$ ($\mathit{3940}$) & $1244$ 
  ($\mathit{1352}$) & $394$ ($\mathit{411}$) & $94$ ($\mathit{94}$) & 
  $21$ ($\mathit{21}$) & $2$ ($\mathit{2}$)  \\ \hline 

  $1$ & $9574$ ($\mathit{11115}$) & $3455$ ($\mathit{3808}$) & $1095$ 
  ($\mathit{1157}$) & $279$ ($\mathit{289}$) & $57$ ($\mathit{57}$) & $2$ 
  ($\mathit{2}$) & $1$ ($\mathit{1}$)  \\ \hline

  $2$ & $8465$ ($\mathit{9365}$) & $2461$ ($\mathit{2586}$) & $593$ 
  ($\mathit{605}$) & $98$ ($\mathit{101}$) & $5$ ($\mathit{5}$) & $2$ 
  ($\mathit{2}$) & -  \\ \hline 

  $4$ & $2847$ ($\mathit{2878}$) & $427$ ($\mathit{431}$) & $35$ 
  ($\mathit{35}$) & - ($\mathit{1}$) & - & - & -  \\ \hline 
\end{tabular}
\caption{The number of host haloes in each mass bin and at each
  `observation redshift' $z_0$. Note that we only consider host haloes
  whose mass never exceeds the mass at $z=z_0$ by more than $10\%$.
  The number in parenthesis is the {\it total} total number of haloes
  in each mass bin, including those that do not meet this criterion.}
\label{tabsum}
\end{table*}

To  construct  merger trees  we  proceed  as  follows: for  each  halo
$M_{z_i}$,  identified  in the  simulation  snapshot corresponding  to
redshift $z_i$, we define the  progenitors at the previous output time
(at $z_{i+1} = z_i + \Delta z$) as being all haloes containing at least
one  particle that  at $z_i$  will  be in  $M_{z_i}$.  The  progenitor
providing the largest  mass contribution to the parent  halo is termed
the {\em main  progenitor}.  Starting from a host  halo at $z=z_0$, we
iterate this procedure back in  time, thus obtaining a complete merger
tree down to the mass resolution fixed at $10$ particles per halo.

Fig.~\ref{treefig} shows  a schematic representation of  a merger tree
to help us define the terminology used throughout the paper.  The {\em
main branch} of  the merger tree is defined as  the branch tracing the
main  progenitor  of  the  main  progenitor  of  the  main  progenitor
etc...  (i.e.   the  branch  connecting   the  dark  grey   haloes  in
Fig.~\ref{treefig}).  We  will use the term {\em  satellites} to refer
to all progenitor haloes accreted  by the main progenitor, and donating
at  least  $50\%$ of  their  mass  to the  host  halo  at $z=z_0$.  In
Fig.~\ref{treefig} these are  the light grey haloes $A,$  $B$, $C$ and
$D$  (as defined  before they  were  accreted onto  the main  branch).
Extending the same  definition to each of these  satellites haloes, we
can go back  in time and trace for each satellite  its own main branch
and  satellites.  For  example,  satellite $D$  has  accreted its  own
satellites $a$, $b$  and $c$ at some redshift  $z_3$, $z_2$ and $z_1$,
respectively.

We stress  that not  all satellites will  survive to $z=z_0$:  many of
them  will  be  destroyed  by  tidal  effects,  encounters  and  other
dynamical    processes.    As   an    example,   satellite    $a$   in
Fig.~\ref{treefig} only  survives until $z= z_2$. For  this reason, in
order  to  identify  substructures  at  $z_0$, at  any  level  of  the
hierarchy, we need to climb the merger trees of all satellites (at any
level of the hierarchy) only up to satellites that will make it to $z=
z_0$, but not further.  In  the example of Fig.~\ref{treefig}, we will
stop climbing the merger tree of  satellite $D$ at $z_3$, which is the
lowest redshift  when any possible substructure present  inside $D$ at
$z=z_3$ has no self-bound counterpart  at $z_0$. We will call $D(z_3)$
a {\em heirless}  satellites.  In the same spirit,  the other heirless
satellites  in Fig.~\ref{treefig}  are  $A(z_4)$, $B(z_3)$,  $C(z_2)$,
$b(z_3)$  and  $c(z_2$).    Their  self-bound  counterparts  at  $z_0$
constitute the substructure population of the host halo, at all levels
of the hierarchy. In what follows we use subhaloes$^{(i)}$ to refer to
the  $i^{\rm th}$  level of  this substructure  hierarchy,  with $i=1$
corresponding to the first order of subhaloes, i.e.  the ones accreted
directly by  the main progenitor. Hence,  in Fig.~\ref{treefig} haloes
$A$, $B$ and $C$ are subhaloes$^{(1)}$,  while $b$ and $c$ make up the
population of subhaloes$^{(2)}$  of the host halo at  $z=z_0$. We will
use  the term  `substructures' to  refer  to subhaloes  of the  entire
hierarchy  ($i=1,2,3,...$). The  mass of  a substructure  at  $z_0$ is
defined by the clump of  particles, originally belonging to a heirless
satellite,  that at  $z=z_0$  are still  self-bound  within the  tidal
radius of that clump.  This  procedure is iterative, because the tidal
radius itself is defined using the self-bound mass of the clump.

To  implement  this algorithm  in  our code,  assume  that  we have  a
simulation  with $N$  snapshots,  and  let $z_i$  be  the redshift  of
snapshot $i$. Let $z_N$ and $z_0$ be the redshifts of the earliest and
final  snapshot,  respectively.    Starting  from  the  population  of
subhaloes identified by SURV at  $z_0$, we proceed along the following
steps:

\begin{enumerate}
\item [$\bullet$]  given a subhalo $s(z_0)$, accreted  as satellite at
$z_i > z_0$, we climb one  step along its own merging history tree and
read the information about its progenitors at $z_{i+1}$;

\item [$\bullet$] for each of these progenitors, we trace the position
of its  particles to  $z_0$, and  see if there  is a  self-bound clump
within its own tidal radius \citep[as defined in][]{tds};

\item [$\bullet$] if at  least two progenitors identified at $z_{i+1}$
(the main  progenitor and another satellite) have  a counterpart clump
$s'(z_0)$ made of at least  $10$ self-bound particles, this means that
$s(z_0)$  has substructures  inside  itself. We  therefore update  the
previous subhalo catalogue by replacing $s(z_0)$ with the counterparts
$s'_1(z_0)$,  $s'_2(z_0)$, ..., of  the survived  progenitors.  Notice
that one  of these progenitors will  always be the  main progenitor of
$s(z_0)$,  so this procedure  will certainly  re-identify the  core of
$s(z_0)$.   Any  other counterpart  $s'_i(z_0)$  will  identify a  new
subhalo at a further level of the substructure hierarchy.

\item [$\bullet$] if, of all progenitors identified at $z_{i+1}$, only
the main progenitor has a  self-bound counterpart at $z_0$, this means
that  $s(z_0)$  has  no   further  levels  of  substructure,  and  the
(satellite)  main  progenitor  of  $s(z_0)$  at $z_i$  is  a  heirless
satellite.  In this  case we  leave the  subhalo $s(z_0)$  as  is, and
proceed to the next one in the catalogue.
\end{enumerate}

After  we have  scrutinized all  subhaloes  once, we  start again  and
consider  the  updated  subhalo  catalogue  (which  now  contains  new
subhaloes at  one further  level of hierarchy).   We repeat  the steps
described above,  skipping however the subhaloes  coming from heirless
satellites, and  split any new  subhalo as described.  We  iterate the
whole procedure: each  time we replace the subhaloes  - which at $z_0$
may be split in two or  more self-bound sub-units - with the sub-units
themselves.   At  some point,  all  subhaloes  at  $z_0$ will  be  the
counterpart  of  some  heirless  satellite.   We  call  the  resulting
population a catalogue of \emph{substructures}.  This catalogue is the
one we use for our investigation.

\begin{figure*}
 \centering   \includegraphics[width=0.9\hsize]{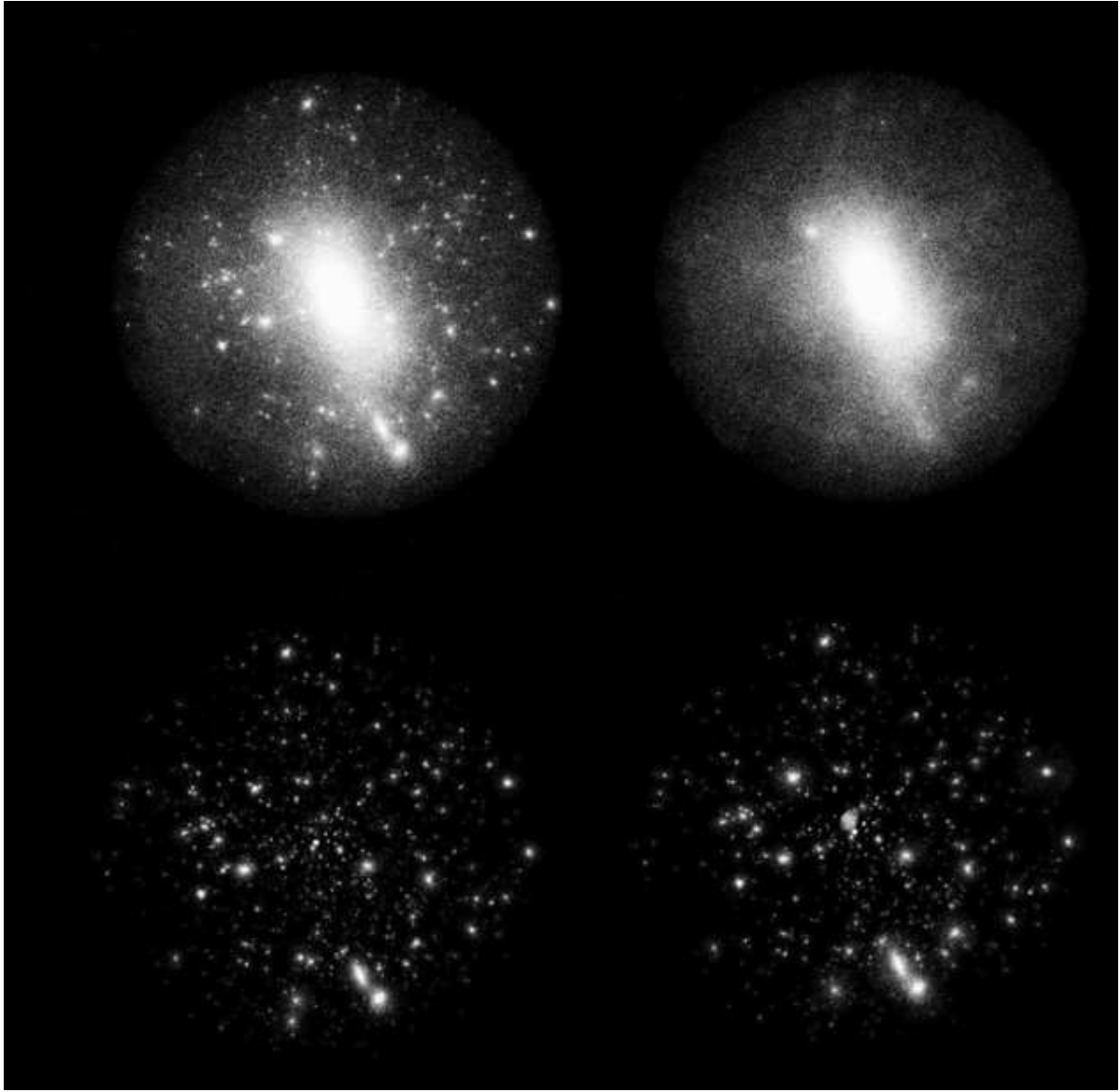}  \vspace{0.5cm}
 \caption{Dark matter particle distributions for the most massive halo
   ($M  = 1.85\times  10^{15}h^{-1}  M_{\odot}$) in  the  GIF2 run  at
   $z=0$, the radius of the sphere is $2.54 h^{-1}$Mpc.  Top left: all
   dark matter particles within the virial radius.  Top right: diffuse
   dark  matter not  associated with  any substructure.   Bottom left:
   substructures,  i.e.  self-bound  particles  belonging to  heirless
   satellites  haloes  (see  the  main  text for  more  details).   The
   particle distribution in the top  right and bottom left sum to give
   that  in  the  top  left.  Bottom  right:  subhaloes$^{(1)}$,  i.e.
   self-bound particles from satellite haloes accreted directly by the
   main progenitor \citep{giocoli08}.  The two bottom panels only show
   substructures and  subhaloes$^{(1)}$ with at  least $10$ particles.
   Clumps at  radii $r < 0.05  R_{\rm vir}$ are not  well defined, and
   were excluded from the study by \citet{giocoli08} and hereafter.}
\label{simfigure}
\end{figure*}

\section{Results}
\label{r2}

Figure~\ref{simfigure} shows the dark matter distribution for the most
massive halo found in the GIF2 run at $z=0$.  The top left panel shows
all particles  inside the  virial radius.  The  top right  panel shows
halo  particles  not bound  to  substructures  (dust particles).   The
bottom left  panel shows particles  bound to substructures (  i.e. the
final result  of our procedure).   The particle distribution  shown in
the top left panel is the sum  of that in this panel and that shown in
the top  right.  The bottom right  panel shows the  particles bound to
subhaloes$^{(1)}$ (i.e. the starting point of our procedure); these are
the objects identified by  \citet{giocoli08}. A comparison between the
two  lower panels shows  that the  largest subhaloes$^{(1)}$  are split
into smaller and smaller substructures as we proceed further along the
merger tree,  until we  reach the heirless  satellites. Note  that the
total  mass   in  the   two  bottom  panels   is  different   for  two
reasons.  First,  some  of  the  particles  in  the  \citet{giocoli08}
subhaloes are  no longer  bound to any  of the  substructures.  Second,
some substructures  found by our  new algorithm (bottom left)  were not
identified  as  subhaloes$^{(1)}$   (bottom  right).   While  the  old
algorithm SURV only follows the  most massive piece, our new algorithm
identifies   all  these   pieces   as  satellites   of  the   original
system. Figure~\ref{massfs} shows the  net change in the mass fraction
associated with substructures.

\begin{figure*}
  \centering \includegraphics[width=16cm]{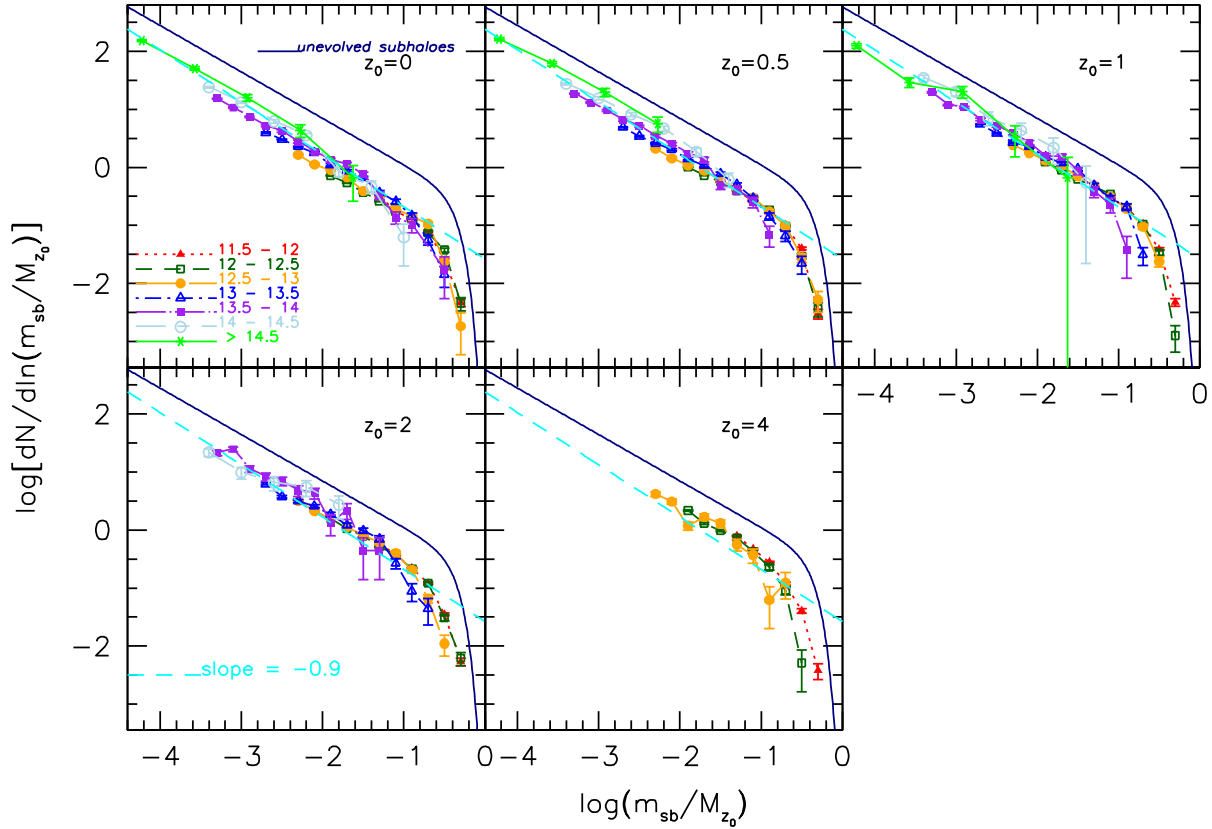} \caption{Substructure
    mass functions at five different redshifts $z_0$ as indicated.  We
    divide the  host haloes in seven  different mass bins  and plot the
    mass function (in  units of the host halo  mass $M_{z_0}$) at that
    redshift. In  each panel  the dashed line  shows a power  law with
    slope  $\alpha=-0.9$ and normalization  $N_0=0.03$ that  fits well
    the  data points  at  $z_0=0$.  For  comparison,  the solid  curve
    represents    the   \emph{unevolved}    subhalo    mass   function
    (equation~\ref{eqsub1}).}
 \label{survf}
\end{figure*}

In  what follows  we discuss  the substructure  of the  host  haloes 
as function of mass and redshift.  We consider five values of 
$z_0$: 0, 0.5, 1, 2 and 4, and for each, we consider seven mass bins. 
Throughout  we only  use host  haloes whose  main progenitor
never exceeds the final mass $M_{z_0}$ by more than $10\%$. 
Table~\ref{tabsum} lists the number of haloes in each of these bins. 
Note that substructures in the innermost region (within $5\%$ of the 
virial radius) are  excluded from further  study, as their 
identification is uncertain due to the extreme density of the host.

The   resulting    substructure   mass   functions    are   shown   in
Figure~\ref{survf}; masses are in units of the virial mass of the host
halo.  The various  symbols and line types in  each panel show results
for different host masses.  The solid curve (same in each panel) shows
the   \emph{unevolved}    subhalo   mass   function   \cite[equation~2
of][]{giocoli08}, i.e., the mass function of progenitor haloes accreted
onto the main branch of the merger tree:
\begin{eqnarray}
  \frac{\mathrm{d}N(m_{\rm sb}|M)}{\mathrm{d} \ln m_{\rm sb}} = 
  N_{0}(M_{z_0}) \, \xi^{\alpha} \,\exp{\left(-\beta\xi^3 \right)} 
  \equiv \xi \, f(\xi) \,, 
\label{eqsub1}
\end{eqnarray}
with  $\xi  = m_{\rm  sb}/M_{z_0}$,  $\alpha=-0.8$,  $\beta =  12.2715
\approx  0.43^{-3}$ and normalization  $N_{0}(M_{z_0}) =  0.18$.  Note
that the  normalization is  independent of both  the mass of  the host
halo  and the  redshift at  which it  was identified  even  though our
notation suggests otherwise.  The  reason for our notation will become
clear shortly.  For comparison, the dashed line shows a power law with
slope $\alpha=-0.9$ and normalization $N_0 = 0.03$: it is steeper than
the solid  line. Performing least squares interpolation  on the linear
trend of the substructure mass  functions we notice that this value of
the slope is in agreement with the value estimated in the more massive
bins, while it tends to overestimate a little the slope in the smaller
ones. This  because subhaloes in  the latter are mainly  preserved and
not split in sub-subhaloes.

\citet{giocoli08}  have shown  that, even  after tidal  stripping, the
evolved mass  function of subhaloes$^{(1)}$ (i.e.  for  the objects in
the bottom  right panel of Figure~\ref{simfigure}) has  the same shape
as the  unevolved one  (e.g., power law  slope $\alpha=-0.8$  at small
$m_{\rm sb}/M_{z_0}$).  I.e., the mass function for objects like those
in  the  bottom right  panel  of  Figure~\ref{simfigure}  is given  by
sliding  the  solid curve  downwards  by  an  amount that  depends  on
$M_{z_0}$  (and  $z_0$).   The  reason  for  this  is  not  completely
straightforward,  since  mass-loss  alone  would  be  associated  with
sliding to  the left, not  down.  As discussed in  \citet{vtg05}, this
happens because the mass loss depends on the time spent by a satellite
within the  potential well of  its host, and  \citet{giocoli08} showed
that the mass function of  objects which accreted before and after the
main progenitor had acquired half the  final mass of the halo are {\em
both}  described   by  equation~\eqref{eqsub1},  but   with  different
normalization   factors   (lower  by   factors   of   0.57  and   0.43
respectively).  Crudely speaking, all  the subclumps accreted prior to
the half-mass time have been erased completely, and all those accreted
later are  still present,  but they have  been stripped.   So, crudely
speaking  the  evolved  mass   function  should  be  given  by  taking
equation~\eqref{eqsub1}, but  with normalization lower by  a factor of
$0.43$, and  then shifting it to  the left.  To  match amplitudes, the
typical shift  is about  a factor of  $3$.  However, because  low mass
objects  typically formed  at  higher redshifts,  their subhaloes  have
suffered more mass loss, so the shift is larger for smaller $M_{z_0}$.
Over the mass range where the mass function resembles a power law, the
leftwards shift  resembles a downwards  shift in the  normalization of
equation~\eqref{eqsub1}, so the number  of subhaloes of a given $m_{\rm
sb}/M_{z_0}$ is larger for larger  $M_{z_0}$.  In the next section, we
show that the scatter around this relation is also directly related to
the halo assembly process.

The symbols  in Figure  \ref{survf} do not  show this  evolved subhalo
mass  function.   Rather, they  show  the  substructure mass  function
associated with our new algorithm  (i.e., for the objects shown in the
bottom  left  panel  of  Figure~\ref{simfigure}).   Notice  that  this
function  has  a  steeper  slope  ($\alpha=-0.9$) than  that  for  the
subhaloes$^{(1)}$ (at least in the most massive host haloes).  However,
equation~\eqref{eqsub1} still provides a  good description of the full
shape, if we simply set $\alpha=-0.9$, keep $\beta = 12.2715$, but now
allow the  normalization $N_{0}(M_{z_0})$ to depend on  the mass (and,
we show later, on $z_0$) of the host halo.

\begin{figure}
  \centering \includegraphics[width=\hsize]{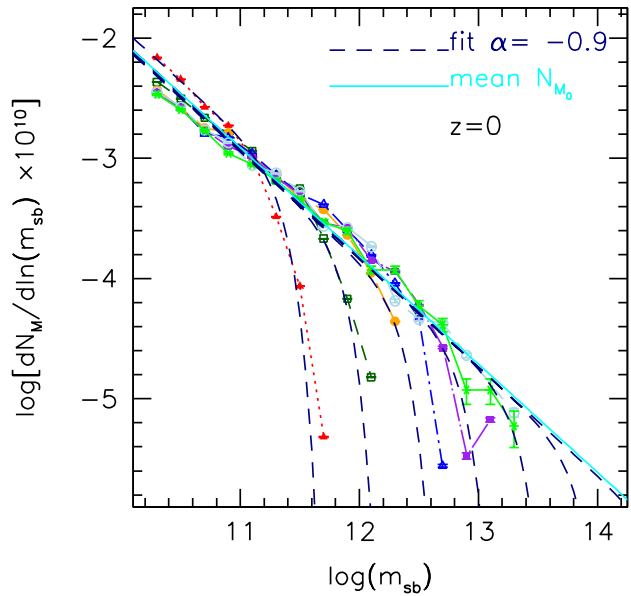}
  \caption{Substructure  mass function  per  unit host  halo mass  (in
    units of $10^{10}\,h^{-1}M_{\odot}$).  Different data points refer
    to the  same mass bins as  in Figure \ref{survf}.   The solid line
    shows a power law distribution with slope $\alpha=-0.9$ and a mean
    normalization computed over the  different mass bins (see the text
    for more details) \label{subuv}. The  dashed lines show the fit to
    the  data  for  each  mass  bin plus  the  high  mass  exponential
    cut-off.}
\end{figure}

\begin{figure}
  \centering  \includegraphics[width=\hsize]{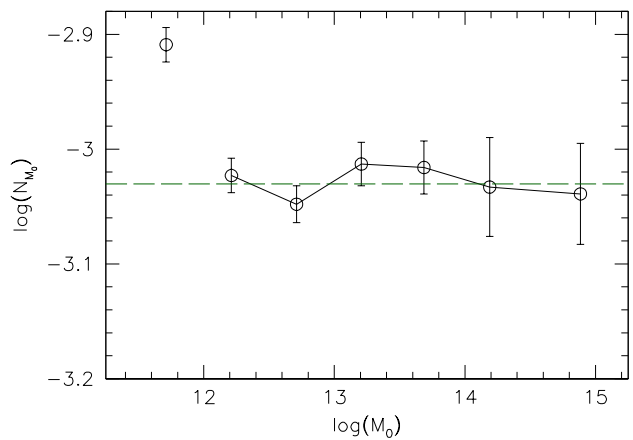} \caption{Host
    halo  mass dependence  of  the normalization  factor $N_{M_0}$  in
    eq. (\ref{substrupm}) at $z_0=0$.  The error bars were obtained by
    bootstrapping the host halo sample  in each mass bin and computing
    the rms of the  normalization factors obtained for the bootstraps.
    The dashed  line in the top  panel shows the mean  weighted by the
    errors of the data points excluding the smallest mass bin (see the
    main text  for more  details).  
    \label{normbut1}}
\end{figure}

Figure \ref{subuv}  shows the substructure  mass function if  one does
not normalize all $m_{\rm sb}$ by $M_{z_0}$.  Namely, it shows
\begin{equation}
  \frac{\mathrm{d}N_M}{\mathrm{d}\ln m_{\rm sb}} \equiv
  \frac{m_{\rm sb}}{M_{0}} 
  \frac{\mathrm{d}N}{\mathrm{d}m_{\rm sb}} 
  = N_{M_0}\, m_{\rm sb}^\alpha\,\exp{\left(-\beta\xi^3 \right)}  ,
\label{substrupm}
\end{equation}
where the normalization factor is
\begin{equation}
  N_{M_0} = \frac{N_{0}(M_{z_0})}{M_0^{1+\alpha}}\,.
\end{equation}
The  dashed curves  show  this  expression with  $\alpha  = -0.9$  and
normalization determined  by chi-square  minimization to the  data for
each  bin  in  $M_0$.   Figure~\ref{normbut1}  shows  that,  with  the
exception  of  the  lowest   mass  bin,  $\log  N_{M_0}=-3.03$  almost
independent of  mass.  We believe the  lowest mass bin,  for which the
haloes  have  the  fewest  particles, is  compromised  by  discreteness
effects: the substructure  mass loss rates of the  objects in this bin
are rather discontinuous.

The  substructure population  in this  GIF2 simulation  has  also been
studied  by \citet{getal04}.   They used  a  different post-processing
pipeline and  their algorithm  to identify substructures,  SUBFIND, is
different  from ours.   So it  is remarkable  that our  mass functions
agree  rather   well:  we  both  find   $\alpha=-0.9$.   However,  our
normalizations  are  different:  they  find  $-3.2$  whereas  we  find
$-3.03$.  This can be traced to the fact that we define haloes as being
$\Delta_{\rm vir}(z)\rho_{\rm crit}(z)$, whereas they fix $\Delta_{\rm
vir} = 200$.  In effect, their haloes are smaller and less massive than
ours  (Appendix~\ref{massris} shows that  the volume  $V_{200} \approx
0.4 \,V_{\rm vir}$  at $z=0$). Hence, at the  same numerical value for
the halo mass, our haloes  are `effectively' more massive than theirs,
which  means  smaller  formation  redshifts,  and  hence  higher  mass
fractions  of   surviving  substructure.    We  believe  this   to  be
responsible          for           the          difference          in
normalization.   Appendix~\ref{previous}   gives   a   more   detailed
discussion  of the  differences  between our  new  algorithm, that  of
\citet{getal04}, and that of \citet{giocoli08}.

\section{Mass fraction and substructures}
\label{r3}

In this section, we show how the substructure mass fraction correlates
with  other properties  of  the halo  (mass, concentration,  formation
time).  Most of these are expected -- our results quantify the trends.
We then study how the  number of substructures depends on the redshift
at which the host halo is identified.

\subsection{Correlations with mass, formation time, and concentration}

The  structure of  a  dark matter  halo  is related  to its  accretion
history  \citep[][and references  therein]{zjmb09}.   To quantify  the
different  halo assembly  histories we  will use  the  formation time,
defined as the  earliest time when the main  progenitor reaches a mass
that  is half  of the  final mass  of the  halo \citep[e.g.,][]{lc93}.
Previous  work has  shown that  massive objects  formed  more recently
\citep[e.g.,][]{lc93,vdB02,st04,giocoli07}.  Systems that form at high
redshift, when  the universe was  denser, are denser.  This,  with the
fact that massive objects formed  more recently gives rise to the mass
concentration  relation  \citep{nfw96,bu01,mac07,neto,zjmb09} and  its
scatter.    In  addition,   using  extended   Press-Schechter  theory,
\citet{vtg05}  showed  that the  unevolved  subhalo  mass function  is
universal,   which  was  confirmed   with  numerical   simulations  by
\citet{giocoli08}.   Hence, they predicted  that less  massive haloes,
which on average form earlier (i.e., accrete their subhaloes earlier),
would have less surviving  substructure at the present, simply because
their substructure  would have been  exposed to tidal stripping  for a
longer  duration. \citet{vtg05} thus  predicted a  correlation between
$f_s$ and  mass, but also that the  scatter in $f_s$ at  fixed mass be
correlated with the formation time of the halo.
\begin{figure}
  \centering  \includegraphics[width=\hsize]{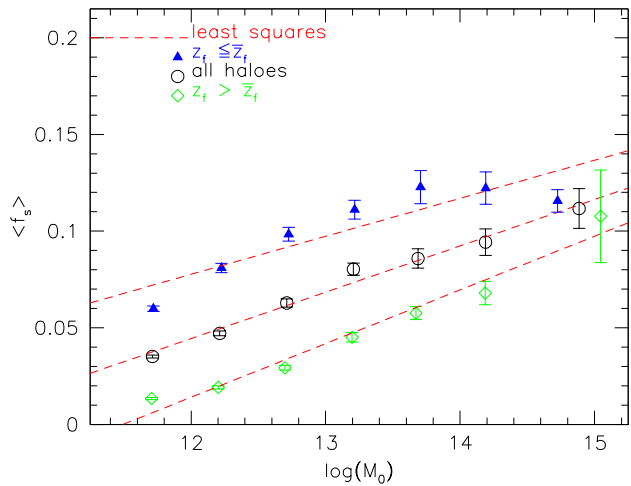}  \caption{Fraction
    of present-day host halo mass in substructures.  Open circles show
    the substructure  mass fraction considering  all haloes in  each of
    the seven mass bins.  Filled triangles and open diamonds represent
    the mean substructure mass  fraction considering only haloes with a
    formation  redshift  lower  and   higher  than  the  mean  in  the
    corresponding mass bin, respectively. The error bar shows the rms.
    The dashed lines show least squares fits to the three sets of data
    points.}
 \label{sufmass}
\end{figure}

To check this, for each halo we computed
\[
 f_s = \frac{\sum_i m_{sb,\,i}}{M_0}\,, 
\]
and, for each mass bin,
\[
\langle      f_s       \rangle      =      \frac{\sum_{j=1}^{N_{haloes}}
  f_{s,\,j}}{N_{haloes}}\,.
\]
When  computing these sums,  we considered  all substructures  with at
least  $10$  self-bound dark  matter  particles  that  were more  than
$0.05R_{\rm vir}$ from the center of the host halo.

Figure~\ref{sufmass}  shows this  mean substructure  mass  fraction at
$z=0$ for a range of host  halo masses.  The open circles show results
for the  full set of  haloes in each  mass bin, whereas  triangles and
diamonds  show  the subset  which  formed  below  and above  the  mean
formation redshift (for  that mass).  At $z=0$, we  have 13467 (7158),
4407 (3898), 1834 (1515), 610 (576), 215 (246), 72 (55), 18 (17) and 2
(2) haloes with a formation redshift higher (lower) than the mean, for
our seven mass  bins.  This shows that the  normalization factor $N_0$
of  the  substructure  mass  function  depends on  $M_0$  and  on  the
formation redshift $z_f$.  For  a given mass, lower formation redshift
translates into a higher substructure mass fraction and smaller smooth
dark  matter component  $M_0 (1-f_s)$.   The three  dashed  lines show
least squares fits to the three sets of data points:
\begin{equation}
\langle f_s \rangle = a \log(M_0) + b\,,
\end{equation}
where $a$ and $b$ for the lines going from the top to bottom in Figure
\ref{sufmass} we have:
\begin{equation*}
 (a,b)= 
 \begin{cases}
  (0.020\pm 0.004, -0.16\pm 0.06) & \text{if $z_f\le\bar{z}_f$,}\\
  (0.024 \pm 0.001, -0.24 \pm 0.02) & \text{for all haloes,} \\
  (0.027 \pm 0.002, -0.32 \pm 0.03) & \text{if $z_f > \bar{z}_f$.}
\end{cases}
\end{equation*}

\begin{figure}
\centering
\includegraphics[width=\hsize]{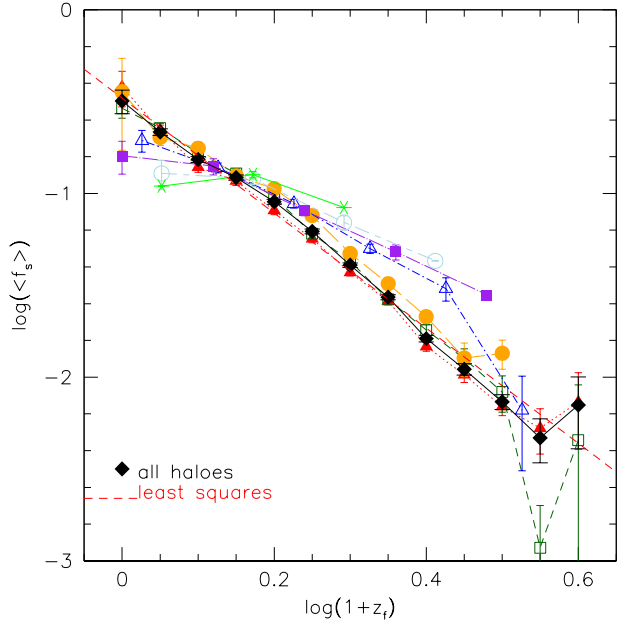}
\caption{Correlation between  the substructure mass  fraction and host
  halo  formation redshift.   Different symbols  show results  for the
  mass  bins  as  in  Figure~\ref{survf}.  Filled  diamonds  show  the
  relation    averaged   over   all    haloes   more    massive   than
  $10^{11.5}M_{\odot}/h$ at  $z=0$; dashed line shows  a least squares
  fit.  \label{fzfc} }
\end{figure}

Figure~\ref{fzfc}   shows  another   way  of   presenting   the  joint
distribution of halo mass, formation time and $f_s$.  In this case, we
study the  correlation between $f_s$  and $z_f$ for  different masses.
The black  diamonds show this  correlation for all haloes  more massive
than  $10^{11.5}M_{\odot}/h$, the  dashed line  shows a  least squares
fit:  
\begin{equation}
 \log\langle f_s \rangle = (-3.133\pm 0.13)\log(1+z_f) - (0.48\pm 0.05),
\end{equation}
and  the other  symbols show  the  correlation for  narrow mass  bins.
Notice that the substructure  mass fraction is tightly correlated with
the halo assembly redshift for  haloes of all masses, indicating that,
to good approximation,  we can write that $f_s  = f_s(z_f)$.  The fact
that this $f_s-z_f$ correlation  depends little on mass $M_0$, whereas
the $f_s-M_0$ correlation depends strongly on $z_f$ indicates that the
$f_s-M_0$ correlation  is almost entirely determined  by the $f_s-z_f$
and $z_f-M_0$ correlations.   From the figure we notice  also that for
haloes formed  quite recently, a  high value of the  substructure mass
fraction indicate that  they have undergone a recent  almost 1:1 major
merging  event.  This  can also  been  understood if  we consider  the
conditional   distribution   of  masses   at   formation  studied   by
\citet{st04a}. In this work they show that the distribution is tightly
peaked around $1/2$ for large formation redshift (since there were few
progenitors of large  mass at high-z), whereas it  is much broader for
lower $z_f$: so low $z_f$ means more recent 1:1 mergers.

\begin{figure}
\centering
\includegraphics[width=\hsize]{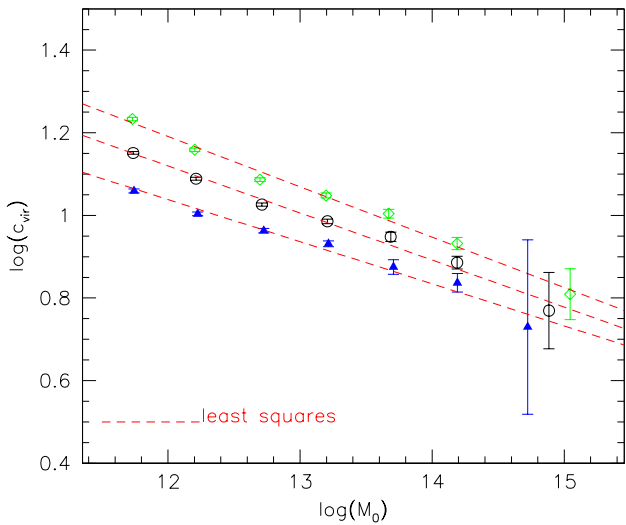}
\caption{Mass--concentration relation at  $z=0$. Open circles show the
  mean for  the full  sample at each  mass; filled triangles  and open
  diamonds show this relation for haloes with formation redshifts that
  are    lower    and   higher    than    the    mean   (similar    to
  Figure~\ref{sufmass}).  Dashed lines show least squares fit to these
  relations.\label{cvirm}}
\end{figure}

\begin{figure}
\centering
\includegraphics[width=\hsize]{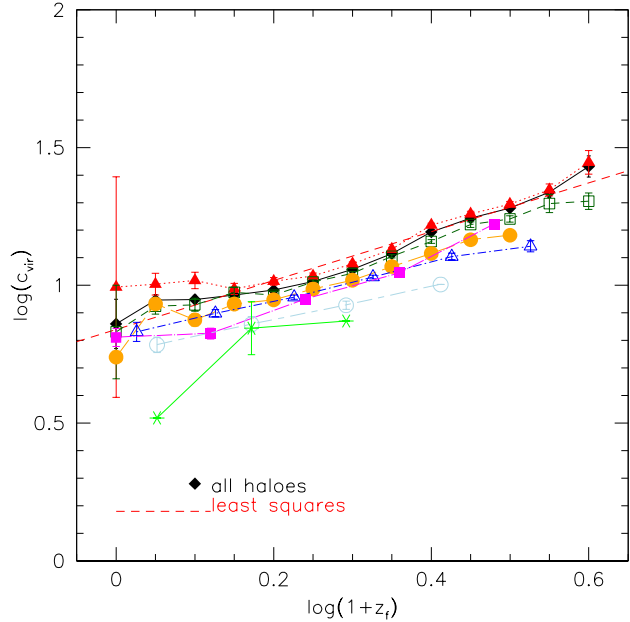}
\caption{Concentration--formation   redshift  correlation.   Different
  symbols and line styles show  results for different mass bins (as in
  Figure~\ref{fzfc}).  Dashed  line shows a  least squares fit  to the
  relation traced by  the filled diamonds, which show  the average over
  all haloes at $z=0$.\label{cvirzf}}
\end{figure}

\begin{figure}
\centering
\includegraphics[width=\hsize]{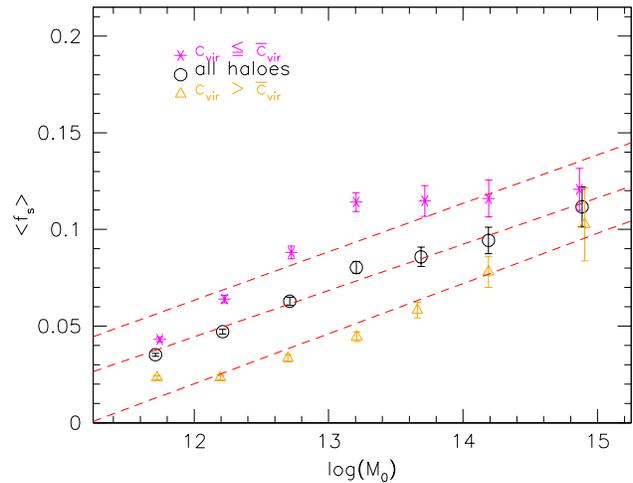}
\caption{Dependence of mass fraction in substructure on host 
  halo mass and concentration.  Open circles show the full sample, 
  triangles and stars show the  average  over systems  with above 
  and  below average concentrations for  the mass bin.  Dashed   
  lines show least squares fits to these trends.  \label{fscm}}
\end{figure}

\begin{figure}
\centering
\includegraphics[width=\hsize]{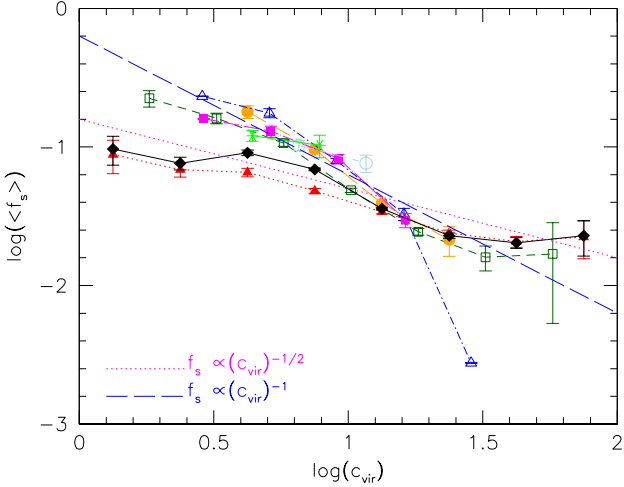}
\caption{Dependence  of mass  fraction  in substructure  on host  halo
  concentration and  mass.  Symbols and line styles  are for different
  host halo masses (same as Figure~\ref{fzfc}).\label{fsmc}}
\end{figure}

Halo formation times  are not observable, so it  is interesting to see
if  halo  substructure correlates  with  other structural  properties.
Halo  concentrations  are potentially  observable,  and  are known  to
correlate        with       mass       and        formation       time
\citep[e.g.][]{nfw97,wetal02,zjmb09}.               Figures~\ref{cvirm}
and~\ref{cvirzf} show these relations in our simulation.  More massive
haloes are less concentrated; at  fixed mass, the objects which formed
at higher redshift are more concentrated.  Different data points refer
to various mass bins, as in previous figures, and error bars represent
the rms  around the  mean value.  We  estimate the  halo concentration
$c_{\rm  vir}=R_{\rm  vir}/r_{\rm s}$  fitting  the spherical  density
distribution  with an NFW  profile.  In  Figure \ref{cvirm}  the three
lines show
\begin{equation}
 \log(c_{vir}) = a \log(M_0) + b
\end{equation}
with 
\begin{equation*}
 (a,b)= 
 \begin{cases}
  (-0.102\pm 0.008,2.26\pm 0.10) & \text{if $z_f\le\bar{z}_f$,}\\
  (-0.114\pm 0.006, 2.49\pm 0.08) & \text{for all haloes,} \\
  (-0.122\pm 0.005, 2.65\pm 0.07) & \text{if $z_f > \bar{z}_f$.}
\end{cases}
\end{equation*}
In Figure \ref{cvirzf} the dashed line shows the least squares fit 
to all masses.  However because the normalization of this correlation 
depends only weakly on host halo mass, it fits the smallest mass 
bin, which dominates the numbers.

Figure~\ref{fscm} shows  the result of  remaking Figure~\ref{sufmass},
but now  replacing host halo formation time  with concentration.  
This shows that, at  fixed halo mass, more concentrated  haloes have 
smaller substructure mass fractions.  The three dashed lines show the 
least squares fits to the data points where 
\begin{equation*}
 (a,b)= 
 \begin{cases}
  (0.251\pm 0.005, -0.24\pm 0.07) & \text{if $c_{vir} \le \bar{c}_{vir}$,}\\
  (0.024\pm 0.001, -0.24\pm 0.02) & \text{for all haloes,} \\
  (0.026\pm 0.003, -0.29\pm 0.03) & \text{if $c_{vir} > \bar{c}_{vir}$.}
\end{cases}
\end{equation*}

\begin{figure*}
  \centering       
 \includegraphics[width=1\hsize]{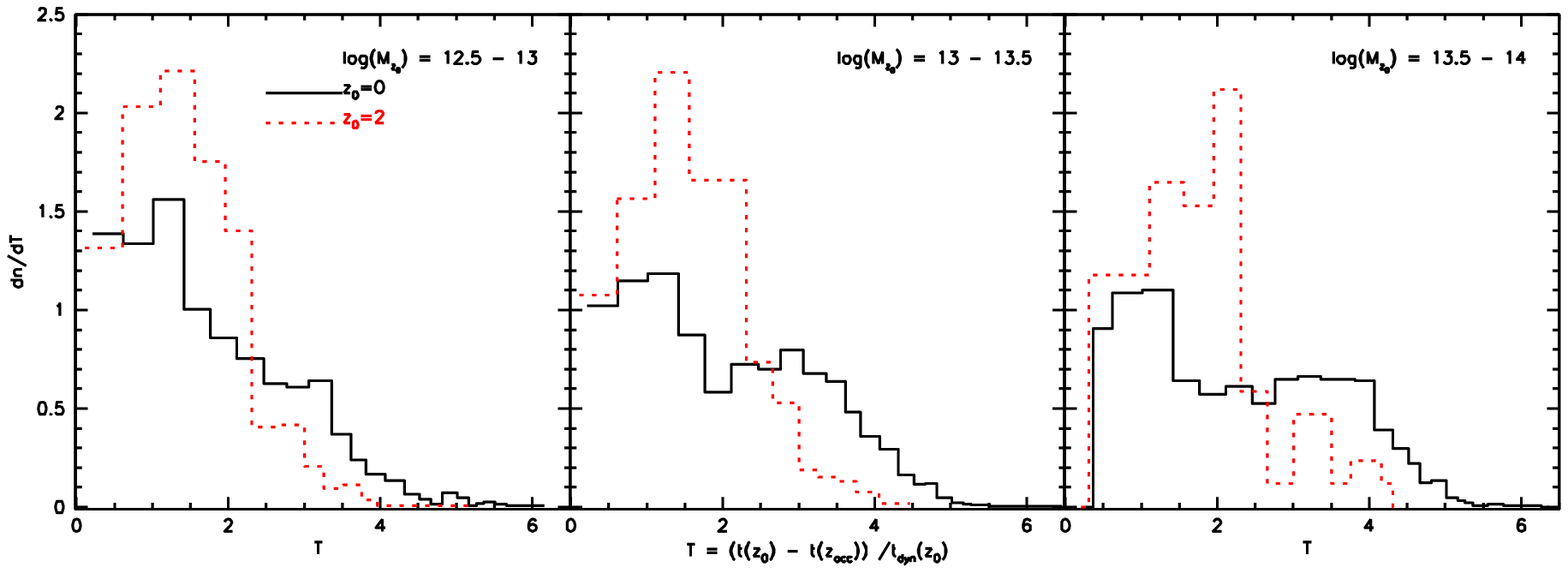}
 \includegraphics[width=1\hsize]{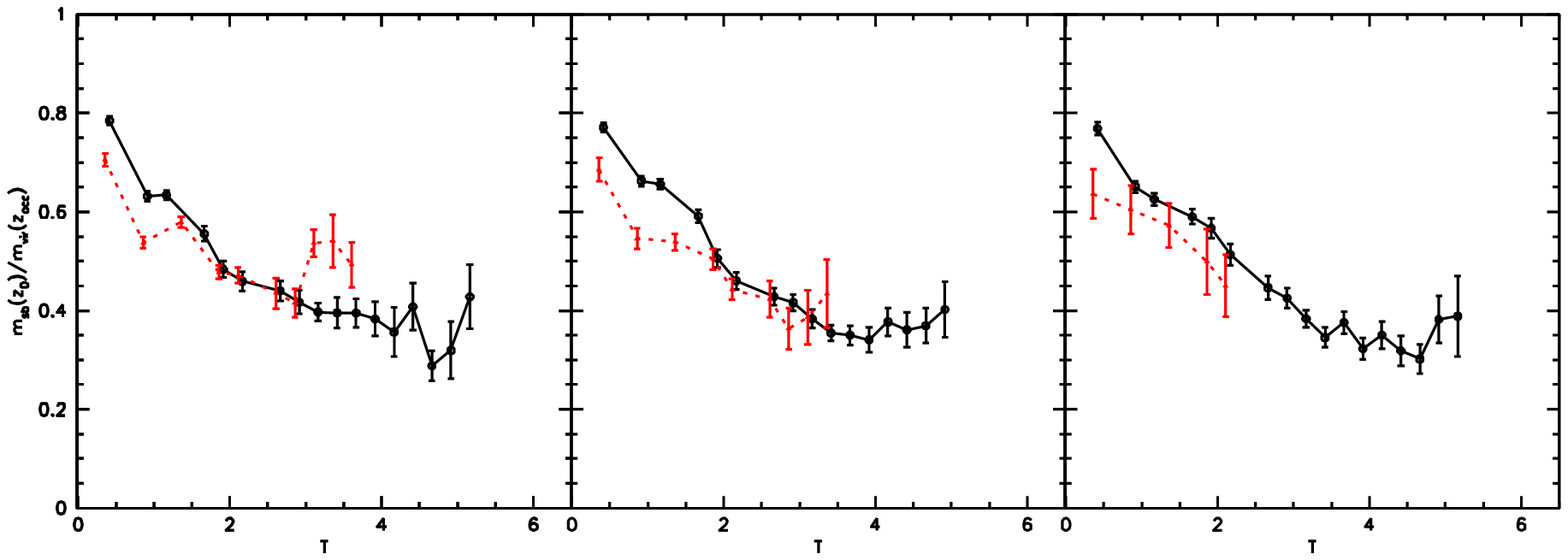}
 \caption{Top panels: Distribution of  the time spent by substructures
   in their host  haloes (i.e. since they were  accreted), in units of
   the dynamical  time scale at $z_0=0$ (solid)  and $z_0=2$ (dashed).
   The three  panels show  results for three  bins in host  halo mass.
   Bottom panels: Retained mass  fraction for substructures which have
   survived within their  host haloes as a function  of (scaled) time,
   for the  same two choices of $z_0$  and for the same  three bins in
   host mass.\label{3frank2}}
\end{figure*}

Figure~\ref{fsmc} shows  a similar remake  of Figure~\ref{fzfc}; $f_s$
decreases  as concentration  increases,  approximately independent  of
halo mass. If we ignore the  smallest mass bin, which is most affected
by  the  mass resolution  of  the  simulation,  then data  points  are
well-fitted by  an inverse correlation  between the mass  fraction and
the concentration of the host halo.

\subsection{Redshift dependence}
At fixed  host halo  mass the number  of substructures depends  on the
redshift  at which the  halo was  identified: it  is larger  at higher
redshift.  Although they did not emphasize this aspect, \citet{angulo}
showed that the subhaloes of hosts identified as being $200$ times the
critical  density in the  Millennium simulation  using SUBFIND  show a
similar trend.

Substructure  abundance depends  on how  much stripping  has occurred,
since this is  what changes the universality of  the unevolved subhalo
population.  This stripping depends on  how long a satellite has spent
inside  its host,  and  on  the dynamical  timescale  within the  host
\citep{vtg05,giocoli08}.   These timescales,  that represent  the time
over which  changes in one part of  a body can be  communicated to the
rest  of  that  body, depend  on  when  the  host halo  is  identified
($\tau_{\rm dyn}\propto  \rho_{vir}^{-1/2})$, so  this may be  why the
higher  redshift hosts  have more  substructures.   If we  use $T$  to
denote the  ratio of  these two timescales  ($T \equiv (t(z)  - t_{\rm
  acc})  /  t_{\rm dyn}(z)$),  then  large  values  of $T$  should  be
associated   with  more   stripping.    The  top   panels  of   Figure
\ref{3frank2}  show  the  distribution  of $T$  for  substructures  in
different  host  halo  masses   (right  to  left)  identified  at  two
redshifts, $z_0 = 0$ and  2 (solid and dashed histograms).  This shows
that, for host haloes observed  at higher redshift the distribution of
$T$ peaks at  lower values.  Substructures in these  systems that have
spent less time in the potential  well of their host suffer less tidal
stripping than  those in  present-day hosts (of  the same  mass).  The
bottom panels show the  correlation between the survived mass fraction
$m_{\rm sb}(z_0)/m_{\rm  vir}(z_{\rm acc})$ and $T$.   As discussed in
\citet{vtg05} and \citet{giocoli08},  when the instantaneous mass loss
rate does  not depend on  the host halo  mass, then the  retained mass
fraction decreases as $\exp{(-T)}$.

\subsection{Poisson model for substructure abundance}

\begin{figure*}
  \centering   \includegraphics[width=16cm]{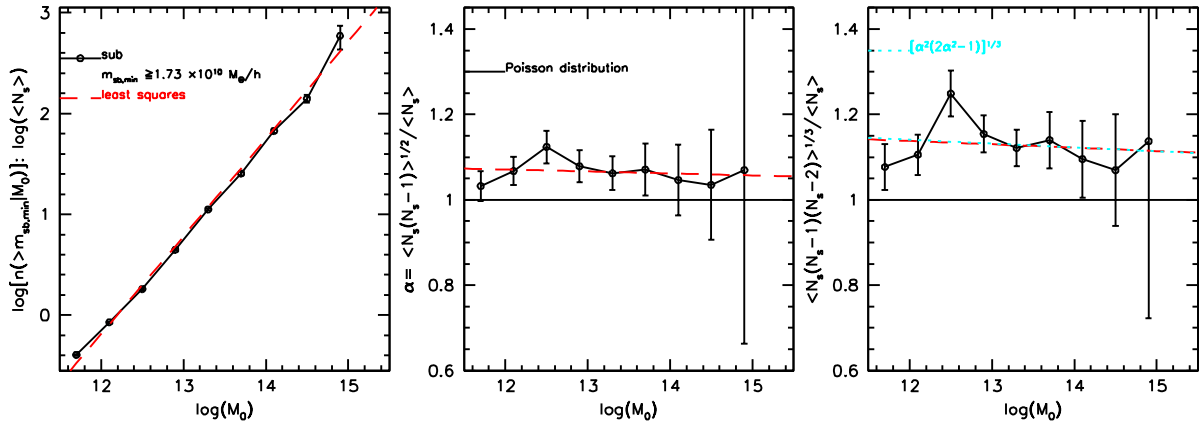}  \caption{First
    three factorial  moments of the substructure counts  as a function
    of  host  halo  mass.   Only  substructures with  more  than  $10$
    self-bound particles,  and further  than $0.05R_{\rm vir}$  of the
    host halo center were  counted.  A Poisson distribution would have
    value unity in the middle and right-most panels.}  \label{hod}
\end{figure*}

Figure  \ref{hod} shows how  the mean  number of  substructures scales
with  host  halo  mass,  again  only counting  substructure  with  ten
particles   or   more  (i.e.,   $m_{\rm   sb,\,min}   =  1.73   \times
10^{10}\,h^{-1}M_{\odot}$).   The  dashed  line  shows the  result  of
fitting
\begin{equation}
\langle N_s \rangle = A_0(m_{\rm sb,\,min}) \, M_0^{\beta},
\end{equation}
to the  measurements.  A least squares  fit returns $\beta  = 0.97 \pm
0.03$   and  $\log(A_0)  =   -11.79  \pm   0.34$.   Of   course,  this
normalization factor  $A_0$ depends on the  smallest substructure mass
considered, but $\beta\approx 1$ is  consistent with the fact that the
normalization  constant  $N_{M_0}$  in  equation~\eqref{substrupm}  is
independent of $M_0$.   The other panels of Figure  \ref{hod} show the
second and third factorial  moments of this distribution: $\langle N_s
(N_s - 1)\rangle$  and $\langle N_s (N_s -  1)(N_s-2)\rangle$.  If the
scatter  around the  mean  were Poisson,  these  would equal  $\langle
N_s\rangle^2$ and $\langle N_s\rangle^3$ respectively.  Evidently, our
substructure counts deviate from the Poisson model only slightly (by a
factor of $\sim  5\%$ at large masses, and  slightly increasing toward
small ones). Interestingly, a similar trend was found by \citet{vtg05}
using merger trees constructed using extended Press-Schechter theory.
\begin{figure}
  \centering \includegraphics[width=\hsize]{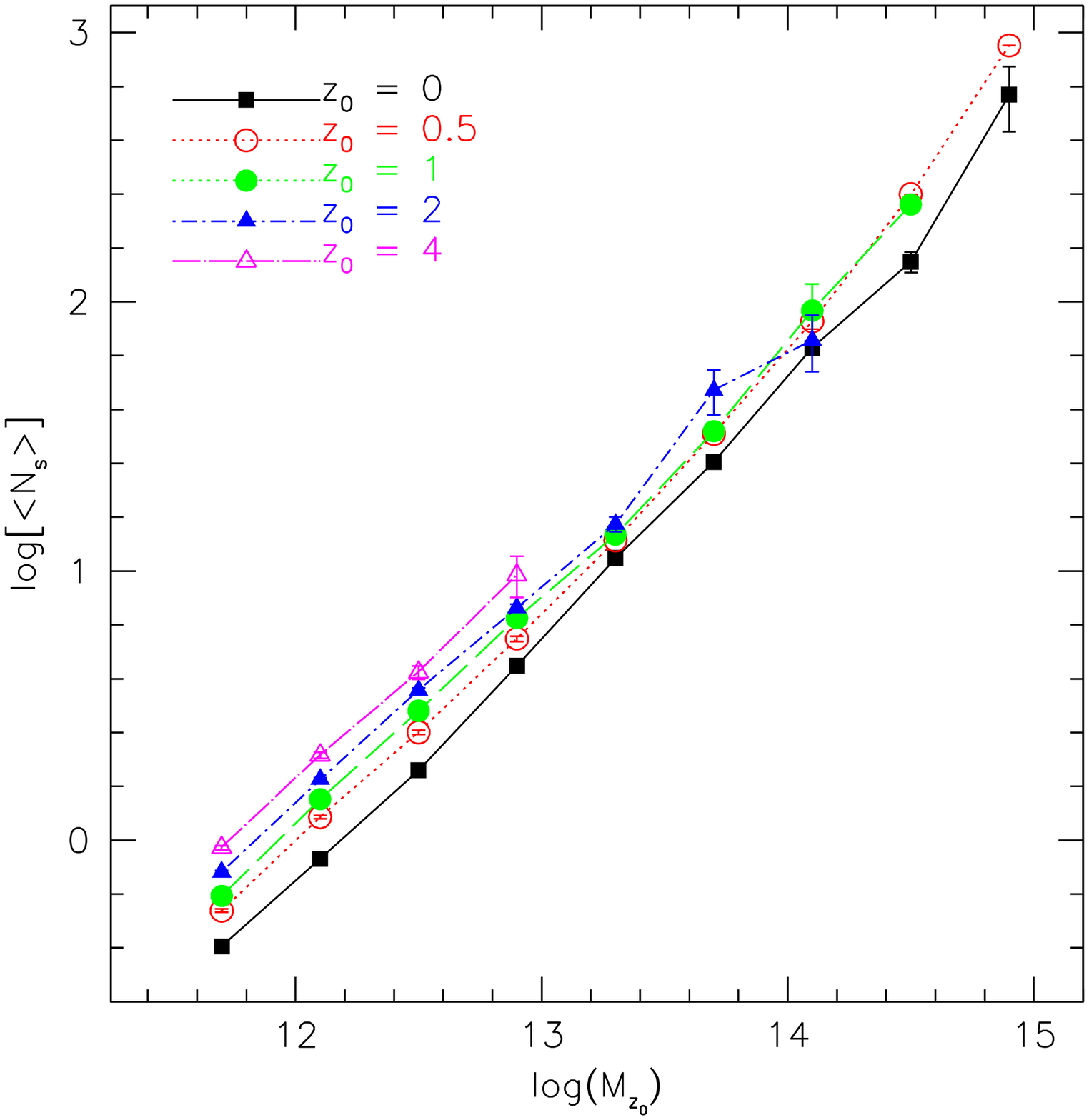} 
  \caption{Dependence of the  mean number of  satellites more massive 
    than  $1.73 \times 10^{10}\,h^{-1}M_{\odot}$  on halo  mass, and  
    on the  redshift at which the  halo was identified. At fixed  mass, 
    host haloes identified at higher redshift have more substructures.  
    \label{zhod}}
\end{figure}

Figure~\ref{zhod}  shows  how  the  mean number  of  satellites  (more
massive  than $1.73\times  10^{10}\,h^{-1}M_{\odot}$)  depends on  the
redshift  at which the  host halo  was identified.   Different symbols
show counts  in haloes at  $z_0 = 0,\,0.5,\,1,\,2,\,4$.   Notice that,
for  the reasons discussed  earlier, at  a given  host mass,  the mean
counts are  higher for  the hosts identified  at higher  redshift, but
that the  curves are  parallel to one  another.  The higher  number of
substructures at high  redshift can also be related  to an increase in
the  galaxy  merger  probability \citep{wetzel09b,hester09}  and  have
implications for galaxy formations  and mergers.  We showed previously
that the substructure mass fraction $f_s$ is anti-correlated with halo
concentration  $c$, approximately  independent  of halo  mass.  For  a
given   host   halo   mass,   $c$   is  smaller   at   high   redshift
\citep[e.g.][]{zjmb09}.  Hence, if the $f_s-c$ anti-correlation is the
same whatever the redshift at  which the host halo is identified, then
one  expects $f_s$  to be  larger  at high  redshifts, in  qualitative
agreement with the trend shown in Figure~\ref{zhod}.

\begin{figure}
\centering
\includegraphics[width=\hsize]{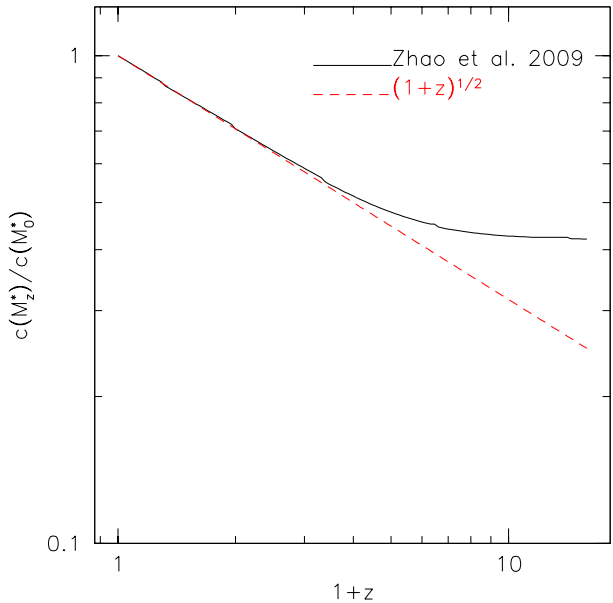}
\caption{Redshift  evolution  of the  concentration  of an  $M^*$-halo
  (here shown in units of the concentration of $M_*$ haloes at $z=0$)  
  as predicted  by the model by \citet{zjmb09}.\label{zhaofig}}
\end{figure}

The  hypothesis  that the  $f_s-c$  anti-correlation  does not  evolve
allows us to  be more quantitative, provided that we  have a model for
how  halo concentrations  depend on  the redshift  at which  they were
identified.   \cite{zjmb09} describe  an accurate  model for  this, in
which a halo's concentration is related  to the time at which its main
progenitor first  assembled $4\%$  of its final  mass.  We  used their
model to  make Figure~\ref{zhaofig}, which shows  the concentration of
an $M_*$ halo as a  function of redshift $z$.  (I.e., $M_*$ represents
the  typical halo  mass  at  redshift $z$,  defined  by $S(M^*_{z})  =
\delta_c(z)^2$,  where  $\delta_c(z)$   is  the  critical  overdensity
required for spherical  collapse at $z$ and $S(M)$  is the variance in
the linear fluctuation field when smoothed with a top-hat filter which
contains mass $M$.  Thus, $M_*$  is smaller at higher redshift.)  This
shows  that,   between  $z=0$  and   4,  $c(M_*(z))/c(M_*(0))  \propto
(1+z)^{-1/2}$: although  $M_*(z)$ is smaller at high  redshift, so too
is its concentration.

\begin{figure}
 \centering
 \includegraphics[width=\hsize]{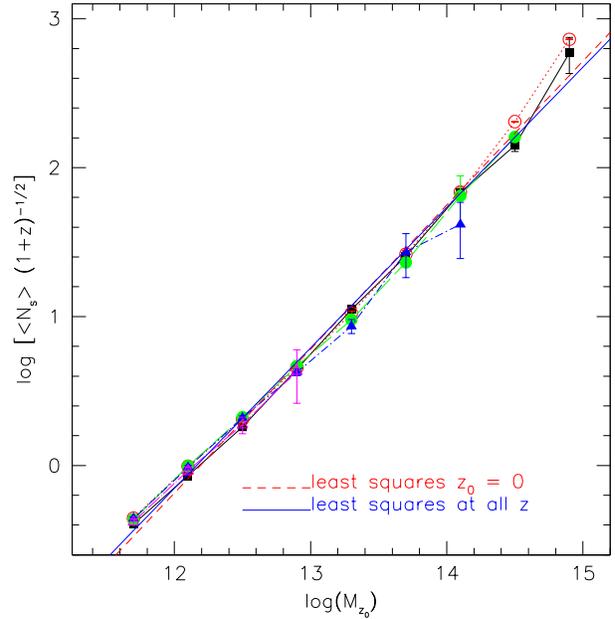}
 \caption{Same as previous figure, but  now rescaled by a factor which
   reflects  the anti-correlation  between  substructure fraction  and
   halo  concentration,  and  the  redshift  dependence  of  the  halo
   concentration.  \label{zhodscaled}}
\end{figure}

\begin{figure}
 \centering
 \includegraphics[width=\hsize]{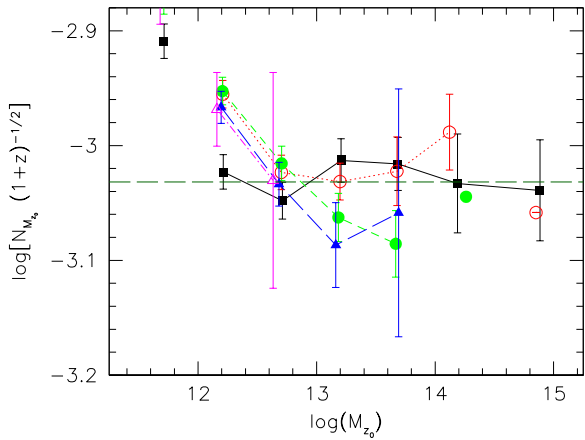}
 \caption{Rescaled   normalization   factor    for   the   number   of
   substructures  per host  halo mass,  shown  in the  same format  as
   Figure~\ref{normbut1}.   Different  symbols   are  for  host  haloes
   identified   at   different   redshifts   $z_0$   (as   in   Figure
   \ref{zhod}).  \label{nvmzz}}
\end{figure}

Since $f_s\propto c^{-1}$ at $z=0$ (Figure~\ref{fsmc}), the hypothesis
that the  $f_s-c$ anti-correlation does  not evolve suggests  that the
typical  value  of  $f_s$  at  some  $z_0$  will  be  proportional  to
$c(M_*(z_0))^{-1}$.   Therefore, subhalo  counts at  $z_0$  should lie
above those  at $z=0$  by a factor  of $[c(M^*_0)/c(M^*_{z_0})]\approx
(1+z_0)^{1/2}$.  Figure~\ref{zhodscaled} shows  the result of dividing
the $z_0$  counts shown in Figure~\ref{zhod}  by $(1+z_0)^{1/2}$: this
does indeed  bring the counts  at different $z_0$ into  good agreement
with   one    another.    To   highlight   how    well   this   works,
Figure~\ref{nvmzz} shows the analogue of Figure~\ref{normbut1} but for
these  rescaled   counts  (symbols  and   line  types  as   in  Figure
\ref{zhod}):  the rescaled  counts are  much  less redshift-dependent,
suggesting that  our crude accounting  for the effects of  the $f_s-c$
correlation on subhalo abundances is reasonably accurate.

\section{Global mass function of substructures}
\begin{figure}
\centering
\includegraphics[width=\hsize]{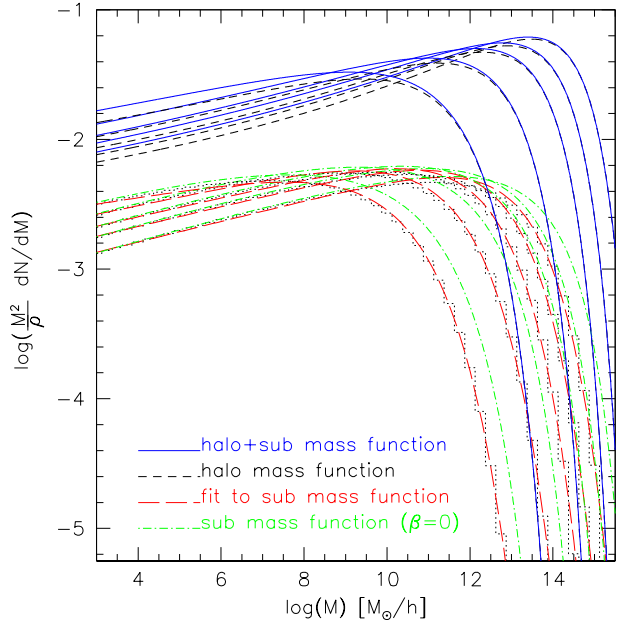}
\caption{Number   density   per   $M^2/\bar{\rho}$  of   substructures
  (histograms) at redshifts  $z=0,\,0.5,\,1,\,2,\,4$ (top to bottom at
  massive  end);   dashed  curves  show  the  result   of  fitting  to
  equation~\eqref{fitsubmf}.  For  comparison, short-dashed curves  show the
  host halo mass function  \citep[from][]{st99} at the same redshifts.
  The  solid line  show the  halo+substructure mass  function  and the
  dot-dashed  the analytical solution  for the  global substructure
  mass         function          when         $\beta\approx         0$
  (eq. \ref{gammamf}).\label{mysub}}
\end{figure}

The mass  function of substructures,  integrated over all  host haloes,
plays  an  important role  in  models  which  relate substructures  to
galaxies \citep{cs02} and to the expected gravitational lensing signal
on   small    scales   \citep{sj03,mm06,nata07,max07}.    The   global
substructure  mass  function can  also  be  useful  for modeling  the
$\gamma$-ray  background  due to  dark  matter particles  annihilation
\citep{forn,gioc08,gioc09}. We  now have the  necessary ingredients to
estimate this quantity:
\begin{eqnarray}
  \frac{\mathrm{d}n(m_{\rm sb})}{\mathrm{d}\ln m_{\rm sb}} &=& 
  \int_{m_{\rm sb}}^{\infty} \frac{\mathrm{d}N(m_{\rm sb}|M)}{\mathrm{d}\ln m_{\rm sb}}
  \frac{\mathrm{d}n(M)}{\mathrm{d}M} \, \mathrm{d}M\nonumber\\
  &=& N_{M_0} (1+z)^{1/2} \, 
  \left(m_{\rm sb}\right)^\alpha\,\nonumber\\
  &&  \times \int_{m_{\rm sb}}^{\infty} 
  \mathrm{d}M\,\frac{\mathrm{d}n(M)}{\mathrm{d} \ln M} \, 
  \exp{\left(-\beta\xi^3 \right)} 
 \label{nusube}
\end{eqnarray}
where $\mathrm{d}n/\mathrm{d}M$ denotes the comoving number density of
dark  matter haloes.   The histograms  in Figure  \ref{mysub}  show the
result  of performing this  integral by  using equation~(\ref{eqsub1})
with $\alpha  = -0.9$, $\beta  = 12.2715$, the  rescaled normalization
from  Figure~\ref{nvmzz} for  $\mathrm{d}N(m|M)/\mathrm{d}m$,  and the
\citet{st99}  expression   for  the   mass  function  of   the  hosts,
$\mathrm{d}n/\mathrm{d}M$, at $z=0,\,0.5,\,1,\,2,\,4$ (top to bottom).
The  short-dashed  curves   show  $\mathrm{d}n/\mathrm{d}M$  at  these  same
redshifts.

\begin{table*}
\centering
\begin{tabular}{|l|c|c|c|c|r|} \hline
  $z$ & $\eta$ & $a_0$ & $\log\left(\bar{m_0}\right)$ &  $\gamma$ &
  $\sigma_{rms}$ \\ \hline \hline
  $0$ &    $0.07930  \pm 0.0002$ & $7.812 \pm 0.001$ & $13.10^{-0.07}_{-0.06}$ & $0.407_{-0.005}^{+0.025}$ & $0.027$ \\ \hline
  $0.5$&   $0.07631  \pm 0.0002$ & $7.944 \pm 0.001$ & $12.61^{-0.06}_{-0.06}$ & $0.386_{-0.005}^{+0.020}$ & $0.026$ \\ \hline
  $1$ &    $0.07308  \pm 0.0002$ & $8.032 \pm 0.001$ & $12.13^{-0.06}_{-0.05}$ & $0.366_{-0.012}^{+0.017}$ & $0.023$ \\ \hline
  $2$ &    $0.06720  \pm 0.0003$ & $8.142 \pm 0.001$ & $11.29^{-0.10}_{-0.05}$ & $0.336_{-0.012}^{+0.018}$ & $0.019$ \\ \hline
  $4$ &    $0.05222  \pm 0.0003$ & $8.267 \pm 0.002$ & $10.06^{-0.08}_{-0.03}$ & $0.309_{-0.008}^{+0.013}$ & $0.016$ \\ \hline
\end{tabular}
\caption{Best fit parameter to the global substructure mass function 
  at the five considered redshifts. The last column shows the rms scatter 
  between the distribution   and the best fit.\label{tabfin}}
\end{table*}

Because  of  the exponential  cutoff  in  the  subhalo mass  function,
equation~\eqref{nusube}  cannot be integrated  analytically.  However,
we        have        found        that       $y(m_{\rm sb})        \equiv
m_{\rm sb}\,\mathrm{d}n(m_{\rm sb})/\mathrm{d}\ln m_{\rm sb}$ is well fit by
\begin{equation}
  y(m_{\rm sb}) = A_0 \, m_{\rm sb}^{\eta}\,
  \exp{\left[ - \left( \frac{m_{\rm sb}}{\bar{m}_0} \right)^{\gamma}
    \right]}. \label{fitsubmf}
\end{equation}
A  similar  functional form  has  been  used  to fit  galaxy  velocity
dispersions \citep{sdssvf}, so our parametrization is intended to help
clarify  the connection  between galaxies  and halo  substructure.  To
determine the  parameters $A_0$, $\eta$, $\gamma$  and $\bar{m}_0$, we
perform a least-squares fit to the log of the counts at the small mass
end: this determines the slope $\eta$ and zero-point $a_0$ of the power
law at  small masses.  We  set $A_0 =  10^{a_0}$, and we  estimate the
other two parameters, $\gamma$ and $\bar{m}_0$, by minimizing
\begin{equation}
  \chi^2 = \sum_i \left[ \log(y(m_{\rm sb})) - 
    \log\left( m_{\rm sb}\frac{\mathrm{d}n(m_{\rm sb})}
      {\mathrm{d}\ln m_{\rm sb}}\right)_i 
  \right]^2  \label{chi} \,,
\end{equation}
where the sum is over the  various values of $\ln m_{\rm sb}$ at which
we  computed the  integral (the  centers  of the  histograms shown  in
Figure~\ref{mysub}).  Table~\ref{tabfin} lists  the resulting best fit
parameters.  As  can be seen  from the last  column of the  Table, the
shape  is fit  to an  accuracy of  about $2\%$.   Solid curves  in the
figure show the halo+substructures mass function.

When  $\beta\approx  0$, equation~(\ref{nusube}) becomes 
\begin{eqnarray}
  \frac{\mathrm{d}n(m_{\rm sb})}{\mathrm{d}\ln m_{\rm sb}} &=&
  N_{M_0} (1+z)^{1/2}\left(m_{\rm sb}\right)^\alpha \bar{\rho} \, \nonumber 
  \frac{A}{\sqrt{\pi}}  \\ &\times & \left[ 0.812 \, 
    \Gamma(0.2,0.353 \nu) + \Gamma(0.5,0.353 \nu) \right] 
  \label{gammamf}
\end{eqnarray}
where $A\approx 0.322$ is the  normalization of the halo mass function
and $\nu  = \delta_{c}(z)^2/S(m)$.  The dot-dashed curves in 
Figure~\ref{mysub} show this approximation.  Clearly, neglecting the 
high-mass exponential cut off is not a good approximation.

\section{Summary and Conclusion}
\label{sc}

We estimated the  mass function of substructures by  following all the
branches (rather than just the main branch) of the merger history tree
of  a halo.   This  distribution, $\mathrm{d}N/\mathrm{d}m_{\rm  sb}$,
depends on  the mass of the  host halo, when the  halo was identified,
and its formation time.  However  when the number of substructures per
host  halo   mass  is  considered,   the  distribution  turns   to  be
``universal'' at  a given redshift,  but simulations with  higher mass
resolution are  required to  determine if this  is true even  when the
host halo mass is smaller than $10^{11}h^{-1}M_\odot$.  This universal
shape is characterized  by a power law of slope  $-0.9$ at low masses,
times  an exponential  cut-off when  the substructure  mass approaches
that of the host halo.   This distribution can be integrated to obtain
the substructure virial mass fraction and the mean number of clumps at
a given host halo mass.

At fixed  host halo mass, the  haloes which formed  at higher redshift
tend to have less substructure.  In addition, more massive host haloes
have  more substructures  than  less massive  ones.   The first  trend
arises because satellites which spend  more time in the potential well
of their host lose a larger  fraction of their initial mass (or can be
totally disrupted).  Massive haloes formed more recently, so this same
mechanism explains  the second trend.   Similar trends are seen  if we
replace  halo  formation  time  (which  is not  observable)  with  the
concentration parameter of the halo density profile (which is).

We also  showed that the  Poisson model, with  a mean that  depends on
host halo mass, is a reasonable description of the substructure counts
within a host  halo.  However, the second and  third factorial moments
of  the substructure counts  are $\sim  5$ percent  larger than  for a
Poisson  model with the  same mean  counts.  As  observations improve,
this  may become  important in  Halo Model  parametrization  of large
scale structure which relate halo substructure to galaxies.

At fixed  mass, the  mean number of  substructures above  some minimum
mass is larger  at high redshift.  Although this  is not unexpected --
the  higher  redshift  haloes  have  had less  time  to  destroy  their
substructures  -- we  argued  that the  redshift  dependence could  be
well-approximated  by  assuming   that  the  anti-correlation  between
substructure abundance and host halo concentration we measure at $z=0$
does  not evolve.   At  fixed  mass, higher  redshift  haloes are  less
concentrated --  in effect,  our assumption is  that the  same physics
which affects the concentrations  also affects the substructure.  This
allows us to provide a  simple prescription for the redshift evolution
of substructures, which we  use to provide analytic approximations for
the result of integrating the substructure mass function over all host
halo   masses.   These  results   should  find   use  in   halo  model
interpretations of the (evolution  of the) galaxy clustering signal in
large scale sky surveys \cite[e.g.,][]{vdb07,2dfsdss}.

\section*{Acknowledgments}

CG  thanks Robert  Smith  and  Jorge Moreno  for  the hospitality  and
suggestions  during his  period  spent in  Philadelphia, and  Matthias
Bartelmann  for useful  discussion  during the  final  writing of  the
paper.   Gao Liang  for providing  us SUBFIND  catalogues of  the GIF2
simulation.    RKS  thanks  R.    Skibba  and   the  members   of  the
MPI-Astronomie, Heidelberg for  their hospitality, and NSF-AST 0908241
for support. 
The authors thank also the referee for useful discussions and comments

\bibliographystyle{mn2e}

\begin{thebibliography}{}

\bibitem[Angulo et al. (2008)]{angulo} Angulo, R.~E., Lacey, 
C.~G., Baugh, C.~M., \& Frenk, C.~S 2008, arXiv:0810.2177 

\bibitem[Benson et al.(2004)]{benson} Benson, A.~J., Lacey, 
C.~G., Frenk, C.~S., Baugh, C.~M., \& Cole, S.\ 2004, MNRAS, 351, 1215 

\bibitem[Bond et al. (1991)]{bond91} Bond, J.~R., Cole, S., Efstathiou,
  G., \& Kaiser, N.\ 1991, ApJ, 379, 440

\bibitem[Brada{\v  c}  et  al. (2004)]{bradacetal04} Brada{\v  c},  M.,
  Schneider, P.,  Lombardi, M., Steinmetz, M.,  Koopmans, L.~V.~E., \&
  Navarro, J.~F.\ 2004, Aap, 423, 797

\bibitem[Bryan \& Norman (1998)]{bn98} Bryan, G.~L., \& Norman, M.~L.\ 1998, ApJ, 495, 80 

\bibitem[Bullock  et al. (2001)]{bu01}  Bullock, J.~S.,  Kolatt, T.~S.,
  Sigad,  Y.,  Somerville,  R.~S.,  Kravtsov,  A.~V.,  Klypin,  A.~A.,
  Primack, J.~R., \& Dekel, A.\ 2001, MNRAS, 321, 559

\bibitem[Cooray \& Sheth (2002)]{cs02} Cooray,  A., \& Sheth, R.\ 2002,
  PhR, 372, 1

\bibitem[Croton  et al. (2006)]{croton}  Croton, D.~J.,  et  al.\ 2006,
  MNRAS, 365, 11

\bibitem[Davis et al. (1985)]{fof} Davis, M., Efstathiou, 
G., Frenk, C.~S., \& White, S.~D.~M.\ 1985, ApJ, 292, 371 

\bibitem[De  Lucia et al. (2004)]{delucia04}  De Lucia,  G., Kauffmann,
  G., Springel, V., White,  S.~D.~M., Lanzoni, B., Stoehr, F., Tormen,
  G., \& Yoshida, N.\ 2004, MNRAS, 348, 333

\bibitem[De Lucia  \& Blaizot (2007)]{db07}  De Lucia, G.,  \& Blaizot,
  J.\ 2007, MNRAS, 375, 2

\bibitem[Diemand et al. (2007)]{diemand07}  Diemand, J., Kuhlen, M., \&
  Madau, P.\ 2007, ApJ, 657, 262

\bibitem[Madau et al. (2008)]{mad08} Madau, P., Diemand, J., 
\& Kuhlen, M.\ 2008, ApJ, 679, 1260 

\bibitem[Dolag  et   al. (2004)]{dolag}  Dolag,  K.,   Bartelmann,  M.,
  Perrotta, F.,  Baccigalupi, C.,  Moscardini, L., Meneghetti,  M., \&
  Tormen, G.\ 2004, A\&A, 416, 853

\bibitem[Eke et  al.  (1996)]{eetal96} Eke, V.~R., Cole,  S., \& Frenk,
  C.~S.\ 1996, MNRAS, 282, 263

\bibitem[Fornengo et al. (2004)]{forn} Fornengo, N., Pieri, 
L., \& Scopel, S.\ 2004, Physical Review D, 70, 103529 

\bibitem[Gao et al. (2004)]{getal04}  Gao,  L.,  White,  S.~D.~M.,
  Jenkins, A., Stoehr, F., \& Springel, V.\ 2004, MNRAS, 355, 819

\bibitem[Gao  et al. (2005)]{gao05}  Gao, L.,  Springel, V.,  \& White,
  S.~D.~M.\ 2005, MNRAS, 363, L66

\bibitem[Gao \& White (2007)]{gao07} Gao, L., \& White, S.~D.~M.\ 2007,
  MNRAS, 377, L5

\bibitem[Gao et  al. (2008)]{gao08} Gao, L., Navarro,  J.~F., Cole, S.,
  Frenk, C.~S.,  White, S.~D.~M., Springel, V., Jenkins,  A., \& Neto,
  A.~F.\ 2008, MNRAS, 387, 536

\bibitem[Ghigna  et  al. (2000)]{ghigna2000}  Ghigna,  S.,  Moore,  B.,
  Governato, F., Lake,  G., Quinn, T., \& Stadel,  J.\ 2000, ApJ, 544,
  616

\bibitem[Gill  et  al. (2004a)]{gill1} Gill,  S.~P.~D.,  Knebe, A.,  \&
  Gibson, B.~K.\ 2004, MNRAS, 351, 399

\bibitem[Gill et al. (2004b)]{gill2} Gill, S.~P.~D., Knebe, A., 
Gibson, B.~K., \& Dopita, M.~A.\ 2004, MNRAS, 351, 410 

\bibitem[Giocoli  et al. (2007)]{giocoli07}  Giocoli,  C., Moreno,  J.,
  Sheth, R.~K., \& Tormen, G.\ 2007, MNRAS, 111

\bibitem[Giocoli et al. (2008a)]{giocoli08}  Giocoli, C., Tormen, G., \&
  van den Bosch, F.~C.\ 2008, MNRAS, 386, 2135

\bibitem[Giocoli et al. (2008b)]{gioc08} Giocoli, C., Pieri, L., 
\& Tormen, G.\ 2008, MNRAS, 387, 689 

\bibitem[Giocoli et al. (2009)]{gioc09} Giocoli, C., Pieri, L., 
Tormen, G., \& Moreno, J.\ 2009, MNRAS, 395, 1620

\bibitem[Harker et al. (2006)]{Hel06} Harker, G., Cole, S., Helly, J.,
  Frenk, C., \& Jenkins, A.\ 2006, MNRAS, 367, 1039

\bibitem[Hester  \&   Tasitsiomi(2009)]{hester09}  Hester,  J.~A.,  \&
  Tasitsiomi, A.\ 2009, arXiv:0902.4489

\bibitem[Ishiyama et al. (2008)]{Ishiy08} Ishiyama, T., 
Fukushige, T., \& Makino, J.\ 2008, PASJ, 60, L13 

\bibitem[Ishiyama et al. (2009)]{Ishiy09} Ishiyama, T., 
Fukushige, T., \& Makino, J.\ 2009, ApJ, 696, 2115 

\bibitem[Jenkins et al. (2001)]{j01}  Jenkins, A., Frenk, C.~S., White,
  S.~D.~M.,  Colberg,  J.~M.,   Cole,  S.,  Evrard,  A.~E.,  Couchman,
  H.~M.~P., \& Yoshida, N.\ 2001, MNRAS, 321, 372

\bibitem[Kamionkowski  \&  Liddle (2000)]{kam00}  Kamionkowski, M.,  \&
  Liddle, A.~R.\ 2000, Physical Review Letters, 84, 4525

\bibitem[Kauffmann  et  al.  (1999)]{kcdw99} Kauffmann,  G.,  Colberg,
  J.~M., Diaferio, A., \& White, S.~D.~M.\ 1999, MNRAS, 303, 188

\bibitem[Kazantzidis et al. (2008)]{kaz1} Kazantzidis, S., 
Bullock, J.~S., Zentner, A.~R., Kravtsov, A.~V., 
\& Moustakas, L.~A.\ 2008, ApJ, 688, 254 

\bibitem[Kazantzidis et al. (2009)]{kaz2} Kazantzidis, S., 
Zentner, A.~R., Kravtsov, A.~V., Bullock, J.~S., 
\& Debattista, V.~P.\ 2009, ApJ, 700, 1896 

\bibitem[Knebe et al. (2001)]{knebe01} Knebe, A., Green, A., 
\& Binney, J.\ 2001, MNRAS, 325, 845 

\bibitem[Koposov et al. (2009)]{koposov09} Koposov, S.~E., Yoo, J., 
Rix, H.-W., Weinberg, D.~H., Macci{\`o}, A.~V., \& 
Escud{\'e}, J.~M.\ 2009, ApJ, 696, 2179

\bibitem[Kravtsov  et al.  (2004)]{ketal04}  Kravtsov, A.~V.,  Gnedin,
  O.~Y., \& Klypin, A.~A.\ 2004, ApJ, 609, 482

\bibitem[Lacey  \& Cole (1993)]{lc93}  Lacey, C.,  \& Cole,  S.\ 1993,
  MNRAS, 262, 627

\bibitem[Li et al. (2008)]{lu08}  Li, Y., Mo, H.~J., \&  Gao, L.\ 2008,
  MNRAS, 389, 1419
  
\bibitem[Li  \& Mo (2009)]{limo} Li, Y., \& Mo, H.\ 2009, arXiv:0908.0301 

\bibitem[Macci{\`o}  \&  Miranda (2006)]{mm06}  Macci{\`o},  A.~V.,  \&
  Miranda, M.\ 2006, MNRAS, 368, 599

\bibitem[Macci{\`o} et  al. (2007)]{mac07} Macci{\`o},  A.~V., Dutton,
  A.~A., van den  Bosch, F.~C., Moore, B., Potter,  D., \& Stadel, J.\
  2007, MNRAS, 378, 55

\bibitem[Macci{\`o} et al. (2009)]{mac09} Macci{\`o}, A.~V., 
Kang, X., \& Moore, B.\ 2009, ApJL, 692, L109 

\bibitem[Meneghetti  et al. (2007)]{max07} Meneghetti,  M., Bartelmann,
  M., Jenkins, A., \& Frenk, C.\ 2007, MNRAS, 381, 171

\bibitem[Moore et al. (1998)]{m98} Moore, B., Governato, F., 
Quinn, T., Stadel, J., \& Lake, G.\ 1998, ApJL, 499, L5 

\bibitem[Moore  et  al. (1999)]{mooreetal99}  Moore,  B.,  Ghigna,  S.,
  Governato, F., Lake, G., Quinn,  T., Stadel, J., \& Tozzi, P.\ 1999,
  ApJ, 524, L19

\bibitem[Navarro et al. (1996)]{nfw96} Navarro, J.~F., Frenk, C.~S., \&
  White, S.~D.~M.\ 1996, ApJ, 462, 563

\bibitem[Navarro et al. (1997)]{nfw97} Navarro, J.~F., Frenk, C.~S., \&
  White, S.~D.~M.\ 1997, ApJ, 490, 493

\bibitem[Natarajan et al. (2007)]{nata07}  Natarajan, P., De Lucia, G.,
  \& Springel, V.\ 2007, MNRAS, 376, 180

\bibitem[Neto et al. (2007)]{neto} Neto, A.~F., et al.\ 2007, 
MNRAS, 381, 1450 

\bibitem[Peebles (1980)]{pee80} Peebles, P.~J.~E.\ 1980, 
Research supported by the National Science Foundation.~Princeton, N.J., 
Princeton University Press, 1980.~435 p.,  

\bibitem[Pieri et al. (2008)]{pbb} Pieri, L., Bertone, G., 
\& Branchini, E.\ 2008, MNRAS, 384, 1627 

\bibitem[Press \& Schechter  (1974)]{ps74} Press, W.~H., \& Schechter,
  P.\ 1974, ApJ, 187, 425

\bibitem[Rasia  et   al. (2004)]{rtm04}  Rasia,  E.,   Tormen,  G.,  \&
  Moscardini, L.\ 2004, MNRAS, 351, 237

\bibitem[Ricotti  \&  Gnedin  (2005)]{rg05}  Ricotti, M.,  \&  Gnedin,
  N.~Y.\ 2005, ApJ, 629, 259

\bibitem[Saro et  al.(2008)]{saro} Saro, A., De Lucia,  G., Dolag, K.,
  \& Borgani, S.\ 2008, MNRAS, 391, 565

\bibitem[Scoccimarro  et  al.  (2001)]{scoccetal01}  Scoccimarro,  R.,
  Sheth, R.~K., Hui, L., \& Jain, B.\ 2001, ApJ, 546, 20

\bibitem[{Seljak   \&  Zaldarriaga   (1996)}]{sz96}  Seljak,   U.,  \&
  Zaldarriaga, M.\ 1996, ApJ, 469, 437

\bibitem[Shaw et al.(2006)]{shaw06} Shaw, L.~D., Weller, J., 
Ostriker, J.~P., \& Bode, P.\ 2006, ApJ, 646, 815 

\bibitem[Shaw et al.(2007)]{shaw07} Shaw, L.~D., Weller, J., Ostriker,
  J.~P., \& Bode, P.\ 2007, ApJ, 659, 1082

\bibitem[Sheth (1998)]{sheth1998} Sheth, R.~K.\ 1998, MNRAS, 300, 
1057 

\bibitem[Sheth  et  al. (2003)]{sdssvf}  Sheth,  R.~K.,  Bernardi,  M.,
  Schechter P. L., et al.\ 2003, ApJ, 594, 225

\bibitem[Sheth  \& Diaferio (2001)]{sd01}  Sheth,  R.~K., \&  Diaferio,
  A.\ 2001, MNRAS, 322, 901

\bibitem[Sheth \&  Jain (2003)]{sj03} Sheth, R.~K., \&  Jain, B.\ 2003,
  MNRAS, 345, 529

\bibitem[Sheth  \&  Tormen  (1999)]{st99}  Sheth,  R.~K.,  \&  Tormen,
  G.\ 1999, MNRAS, 308, 119

\bibitem[Sheth  et  al. (2001)]{smt01}  Sheth,  R.~K.,  Mo,  H.~J.,  \&
  Tormen, G.\ 2001, MNRAS, 323, 1

\bibitem[Sheth  \& Tormen (2002)]{st02}  Sheth, R.~K.,  \&  Tormen, G.\
  2002, MNRAS, 329, 61

\bibitem[Sheth  \& Tormen(2004)]{st04a} Sheth,  R.~K., \&  Tormen, G.\
  2004, MNRAS, 349, 1464

\bibitem[Sheth  \& Tormen (2004)]{st04}  Sheth, R.~K.,  \&  Tormen, G.\
  2004, MNRAS, 350, 1385

\bibitem[Simon   \&  Geha (2007)]{simonmw}   Simon,  J.~D.,   \&  Geha,
  M.\ 2007, ApJ, 670, 313

\bibitem[Spergel   et  al.   (2003)]{spetal03}   Spergel,  D.~N.,   et
  al.\ 2003, ApJS, 148, 175

\bibitem[Springel   et  al. (2001)]{swtk01}  Springel,   V.,  White,
  S.~D.~M., Tormen, G., \& Kauffmann, G.\ 2001, NMRAS, 328, 726

\bibitem[Springel et al. (2005)]{speal05}  Springel, V., et al.\ 2005,
  Nat, 435, 629

\bibitem[Springel et al.(2008a)]{acqua01} Springel, V., et al.\ 
2008, MNRAS, 391, 1685 

\bibitem[Springel et al. (2008b)]{acqua02} Springel, V., et al.\ 
2008, Nat, 456, 73 

\bibitem[Stoehr (1999)]{stth} Stoehr,  F.\ 1999,  Diploma  Thesis, at
  Physics                     Department                    Technishce
  Universit$\ddot{\mathrm{a}}$t. M$\ddot{\mathrm{u}}$nchen

\bibitem[Stoehr et  al. (2003)]{setal03} Stoehr,  F., White, S.~D.~M.,
  Springel, V., Tormen, G., \& Yoshida, N.\ 2003, MNRAS, 345, 1313

\bibitem[Susa \& Umemura (2004)]{susa} Susa,  H., \& Umemura, M.\ 2004,
  ApJl, 610, L5

\bibitem[Tormen et  al. (1997)]{tbw97} Tormen, G.,  Bouchet, F.~R., \&
  White, S.~D.~M.\ 1997, MNRAS, 286, 865

\bibitem[Tormen et al. (1998)]{tds} Tormen, G., Diaferio, 
A., \& Syer, D.\ 1998, MNRAS, 299, 728 

\bibitem[Tormen (1998)]{t98} Tormen, G.\ 1998, MNRAS, 297, 648

\bibitem[Tormen  et al. (2004)]{tmy04}Tormen,  G., Moscardini,  L., \&
  Yoshida, N.\ 2004, MNRAS, 350, 1397

\bibitem[van den Bosch (2002)]{vdB02} van den Bosch, F.~C.\ 2002, MNRAS,
  331, 98

\bibitem[van  den Bosch et  al. (2005)]{vtg05}  van den  Bosch, F.~C.,
  Tormen, G., \& Giocoli, C.\ 2005, MNRAS, 359, 1029

\bibitem[van  den Bosch et  al. (2007)]{vdb07}  van den Bosch, F.~C.,
et al.\ 2007, MNRAS, 376, 841

\bibitem[Yoshida, Sheth \& Diaferio (2001)]{ysd01} Yoshida, N., Sheth,
  R.~K., \& Diaferio, A.\ 2001, MNRAS, 328, 669

\bibitem[Walker  et  al. (2009)]{walk}  Walker, M.~G.,  Belokurov,  V.,
  Evans, N.~W., Irwin, M.~J., Mateo, M., Olszewski, E.~W., \& Gilmore,
  G.\ 2009, ApJL, 694, L144

\bibitem[Wake et al. (2008)]{2dfsdss}
  Wake D. A., et al., 2008, MNRAS, 387, 1045

\bibitem[Wechsler et  al. (2002)]{wetal02} Wechsler,  R.~H., Bullock,
  J.~S.,  Primack, J.~R., Kravtsov,  A.~V., \&  Dekel, A.\  2002, ApJ,
  568, 52

\bibitem[Weller et al.(2005)]{weller05} Weller, J., Ostriker, 
J.~P., Bode, P., \& Shaw, L.\ 2005, MNRAS, 364, 823 

\bibitem[Wetzel et al.(2009a)]{wetzel09} Wetzel, A.~R., Cohn, J.~D., \&
  White, M.\ 2009, MNRAS, 395, 1376

\bibitem[Wetzel et al.(2009b)]{wetzel09b}  Wetzel, A.~R., Cohn, J.~D.,
  \& White, M.\ 2009, MNRAS, 394, 2182

\bibitem[White \& Rees (1978)]{wr78}
 White S.~D.~M., Rees M.~J., 1978, MNRAS, 183, 341

\bibitem[Zhao et al. (2009)]{zjmb09}
 Zhao D. H., Jing Y. P., Mo H. J., B\"orner G., 2009, ApJ, submitted
 (arXiv:0811.0828)

\end{thebibliography}

\appendix

\section{Comparison with previous work}\label{previous}

Our  new   substructure  algorithm  is  a  modification   of  that  in
\cite{giocoli08} and is rather different from that of \citet{getal04}.

\begin{figure}
\centering
\includegraphics[width=\hsize]{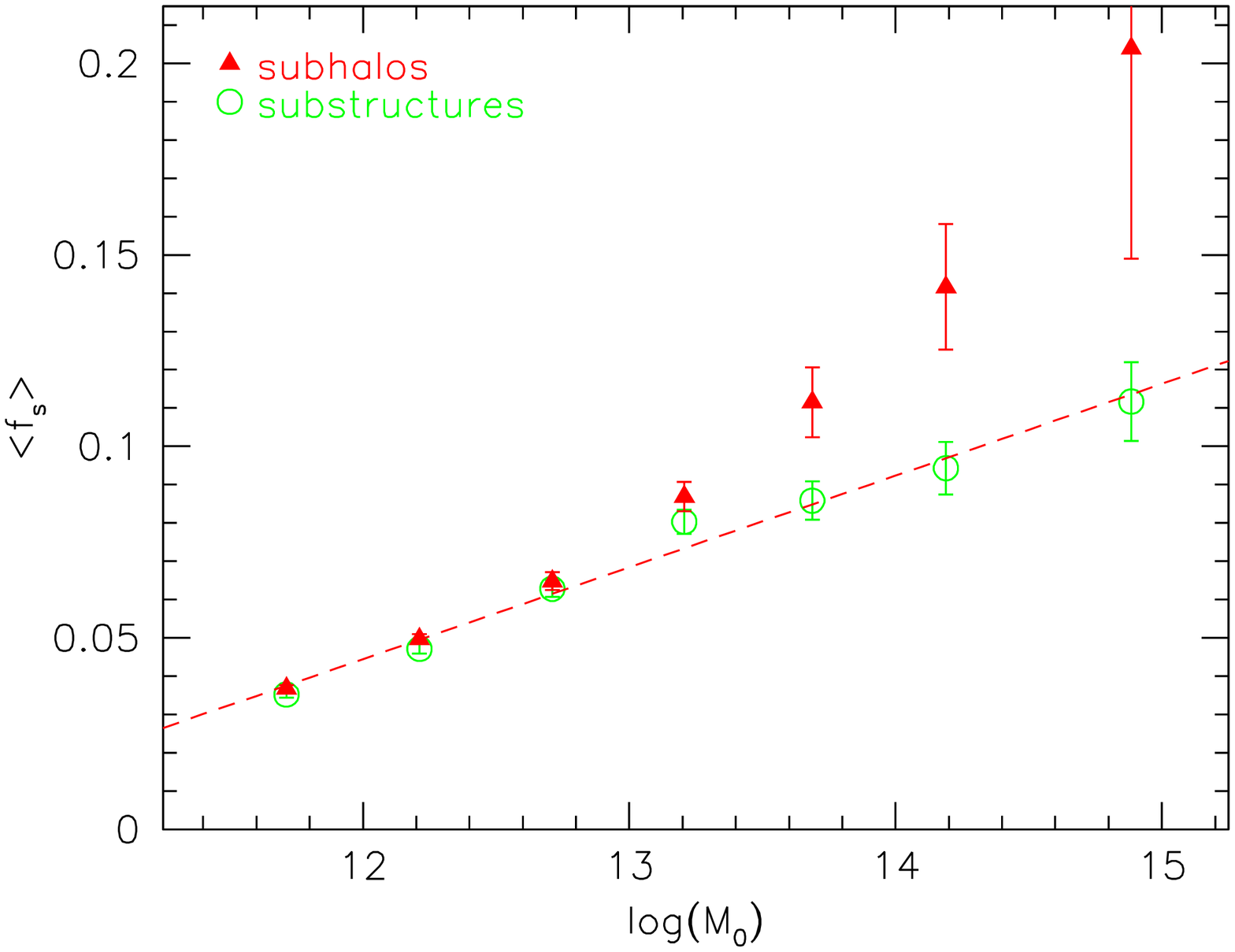}
\caption{Mass  fraction in  the subhaloes  of \citet{giocoli08}  and in
  substructures found  by our new algorithm (filled  and open symbols,
  respectively). \label{massfs}}
\end{figure}

Our  new  algorithm   returns  smaller  substructure  mass  fractions,
especially  at  large   host  halo  masses,  than  did   our  old  one
\citep[i.e.,  that of][]{giocoli08});  it is  more  conservative about
identifying    self-bound    structures.     This    is    shown    in
Figure~\ref{massfs}, where  the change  at the high  mass end  is more
than $50\%$.  It  is smaller at the low mass end,  where the host halo
masses are approaching  that of the mass resolution  of the subhaloes.
This shows explicitly that if one  wants to think of our new algorithm
as taking the subhaloes  of \citet{giocoli08} and partitioning them up
into smaller substructures, then some of the subhalo mass from the old
algorithm  is   now  assigned  to   the  host  halo  rather   than  to
substructures.  The  new value of  the substructure mass  fraction for
the most massive haloes agrees pretty well with other algorithms, that
give values around $10\%$ of the host halo mass\footnote{Comparing our
  new algorithm with  the result of SUBFIND for  the most massive halo
  at $z=0$ we found respectively $6.5\%$ and $6.2\%$ of $M_{200,c}$ to
  be in substructures.}  \citep{swtk01,weller05}.

The  changes in  mass fraction  between  the two  algorithms shown  in
Figure \ref{massfs}  can be understood also considering  the fact that
when  we split  a  subhalo  in pieces  (sub-subhaloes),  we must  also
recompute the  centre-of-velocity (from  which we compute  the kinetic
energy) and,  more importantly, we calculate the  new potential energy
only  using  the  new  subsets  of particles.   This  forces  the  new
potential  energy to  be less  negative  than before  (i.e.  the  same
particle  is assigned a  less negative  potential energy  than before,
because  the potential  wells are  shallower), and  so  many particles
which   reside   in-between  the   density   peaks   of  the   various
sub-subhaloes,  even if they  were bound  to the  subhalo as  a whole,
happen  to be  unbound  to any  of  the sub-subhaloes,  and often  are
unbound to  the main  subhalo itself.  This  is because  the splitting
removed all the secondary density  peaks, so the global potential well
becomes shallower decreasing the self-bound mass of the clumps.

\begin{figure}
\centering
\includegraphics[width=\hsize]{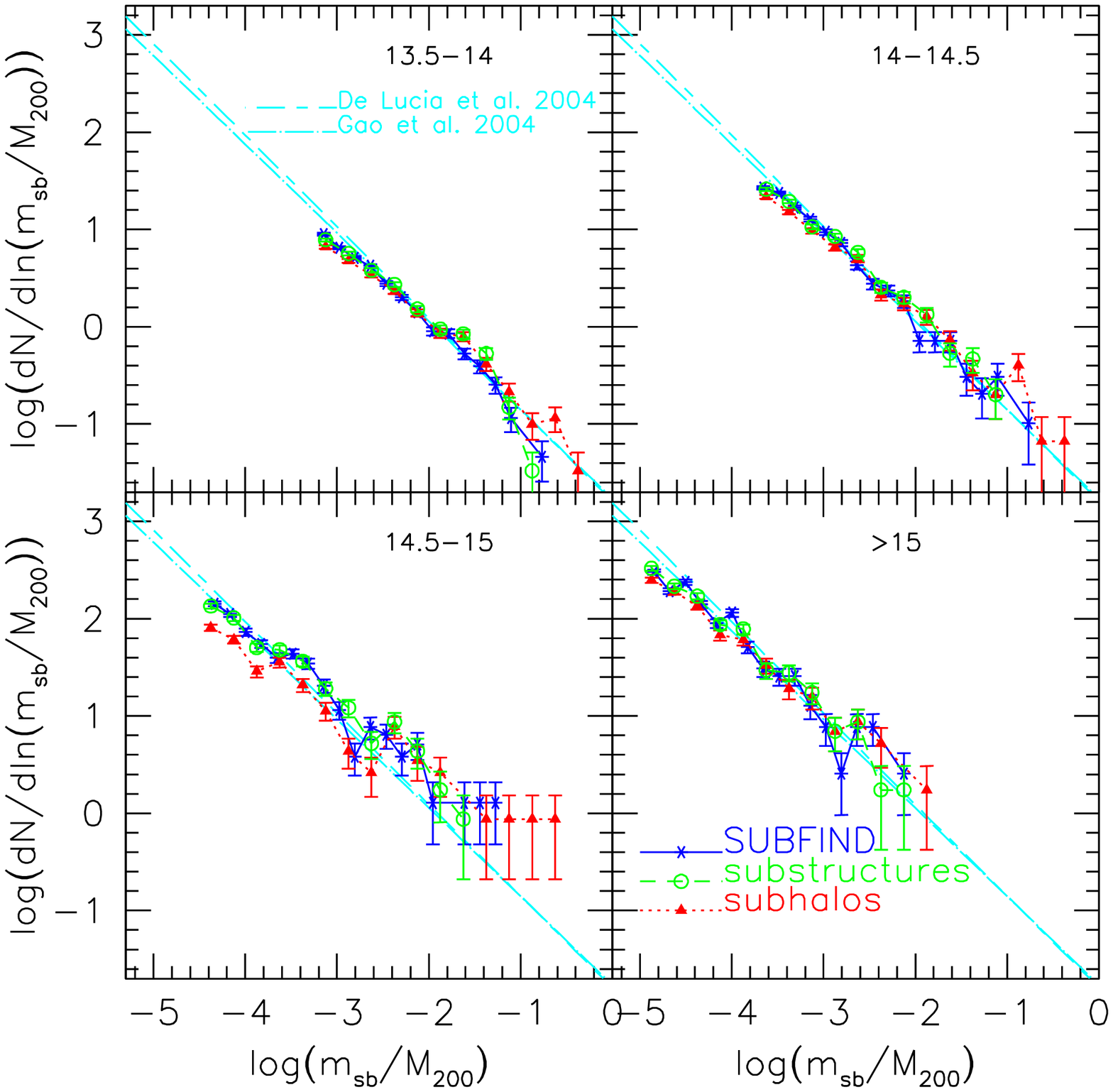}
\caption{Comparison  of substructure  mass functions,  expressed  as a
  function  of $m_{\rm sb}/M_{200}$  found by  our algorithm,  that of
  \citet{getal04}    (SUBFIND),   and    the    subhaloes$^{(1)}$   of
  \citet{giocoli08}.\label{all3}}
\end{figure}

Figure~\ref{all3}  compares the substructure  mass functions  found by
our algorithm, that of  \citet{getal04}, and the subhalo mass function
of \cite{giocoli08}.   To facilitate comparison  with \citet{getal04},
we have chosen  to use the same mass bins and  host halo definition as
\citet{getal04};  i.e.,  host  haloes  are  $200$  times  the  critical
density.  We  will use $M_{200}$  do distinguish this  definition from
$M_{\rm  vir}$,  the  quantity  used  in the  main  text.   Note  that
$M_{200}<M_{\rm vir}$,  with the  difference increasing at  late times
(lower  redshifts).   Our  new  algorithm  (open  symbols)  takes  the
subhaloes returned by our old algorithm (triangles) and partitions them
up into smaller pieces; it  removes objects from large $m_{\rm sb}\sim
M_{200}$ and  adds objects to $m_{\rm sb}\ll  M_{200}$.  The resulting
distribution is  indeed very similar to that  found by \citet{getal04}
using SUBFIND (crosses). The dashed  and the dot-dashed lines show the
fit-by-eye  of   the  cluster  size  halo   by  \citet{delucia04}  and
\citet{getal04}.

\begin{figure}
\centering
\includegraphics[width=\hsize]{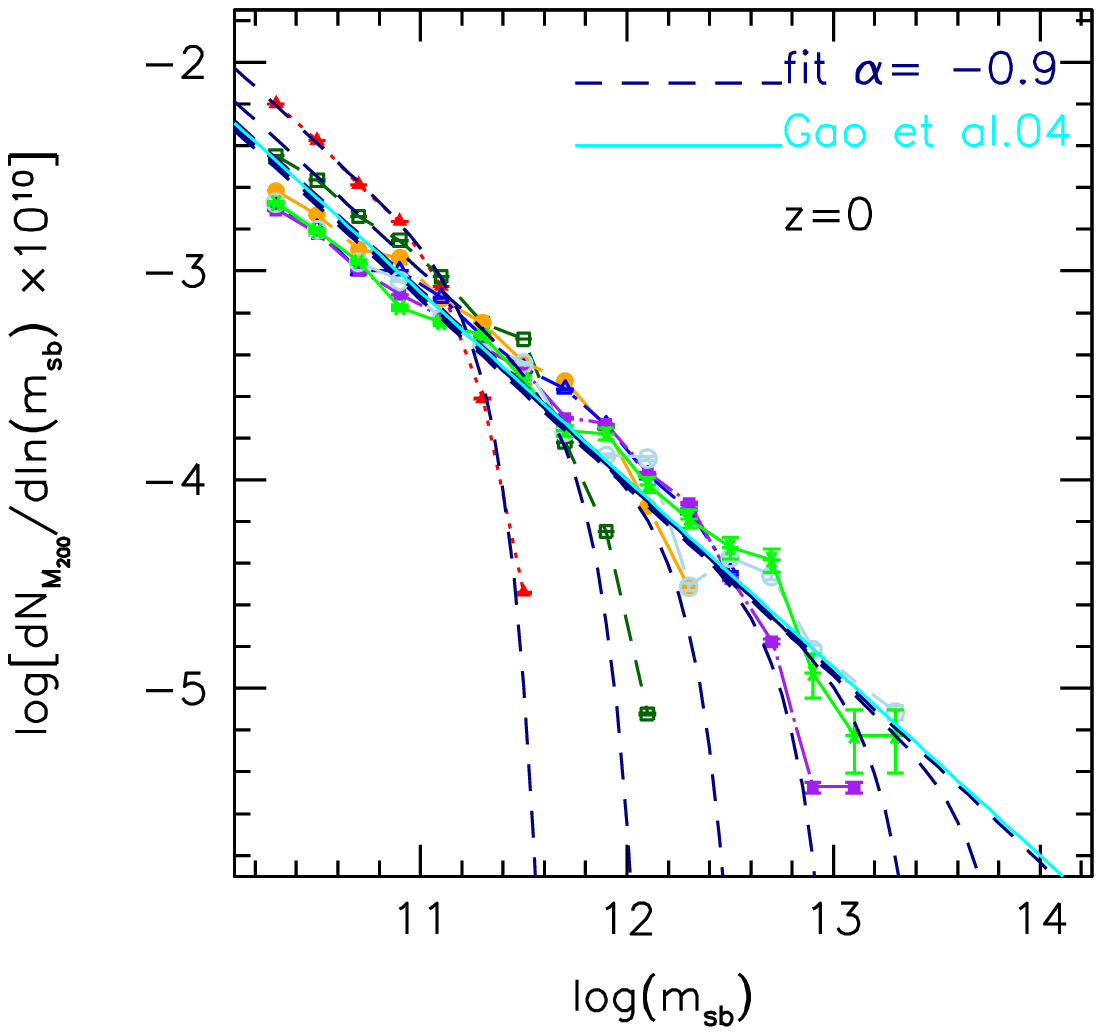}
\caption{Substructure mass functions found  by our algorithm when host
  haloes are defined to be 200 times the critical density.  This is the
  analog  of Figure~\ref{subuv}  in the  main text.   Solid  line show
  equation (1) by \citet{getal04}.  \label{all3perM200}}
\end{figure}

Figure~\ref{all3},      which     shows     $\mathrm{d}N/\mathrm{d}\ln
(m_{\rm sb}/M_{200})$,  is the  analog of  Figure~\ref{survf} in  the main
text.   However,  in  our  discussion  of  $(\mathrm{d}N/\mathrm{d}\ln
m_{\rm sb})/M_{\rm vir}$ in the main text,  we found that the normalization of
our  substructure  mass  function  appeared  to be  larger  that  that
reported  by  \citet{getal04},  and  we  argued  that  this  could  be
attributed to the difference in how we define host halo masses.  Since
the  definitions  are the  same  here, Figure~\ref{all3perM200}  shows
$(\mathrm{d}N/\mathrm{d}\ln m_{\rm sb})/M_{200}$,  which is the  analog of
Figure~\ref{subuv}.  Note  that the normalization is  indeed lower for
$M_{200}$ than it is for $M_{\rm vir}$.

\begin{figure}
\centering
\includegraphics[width=\hsize]{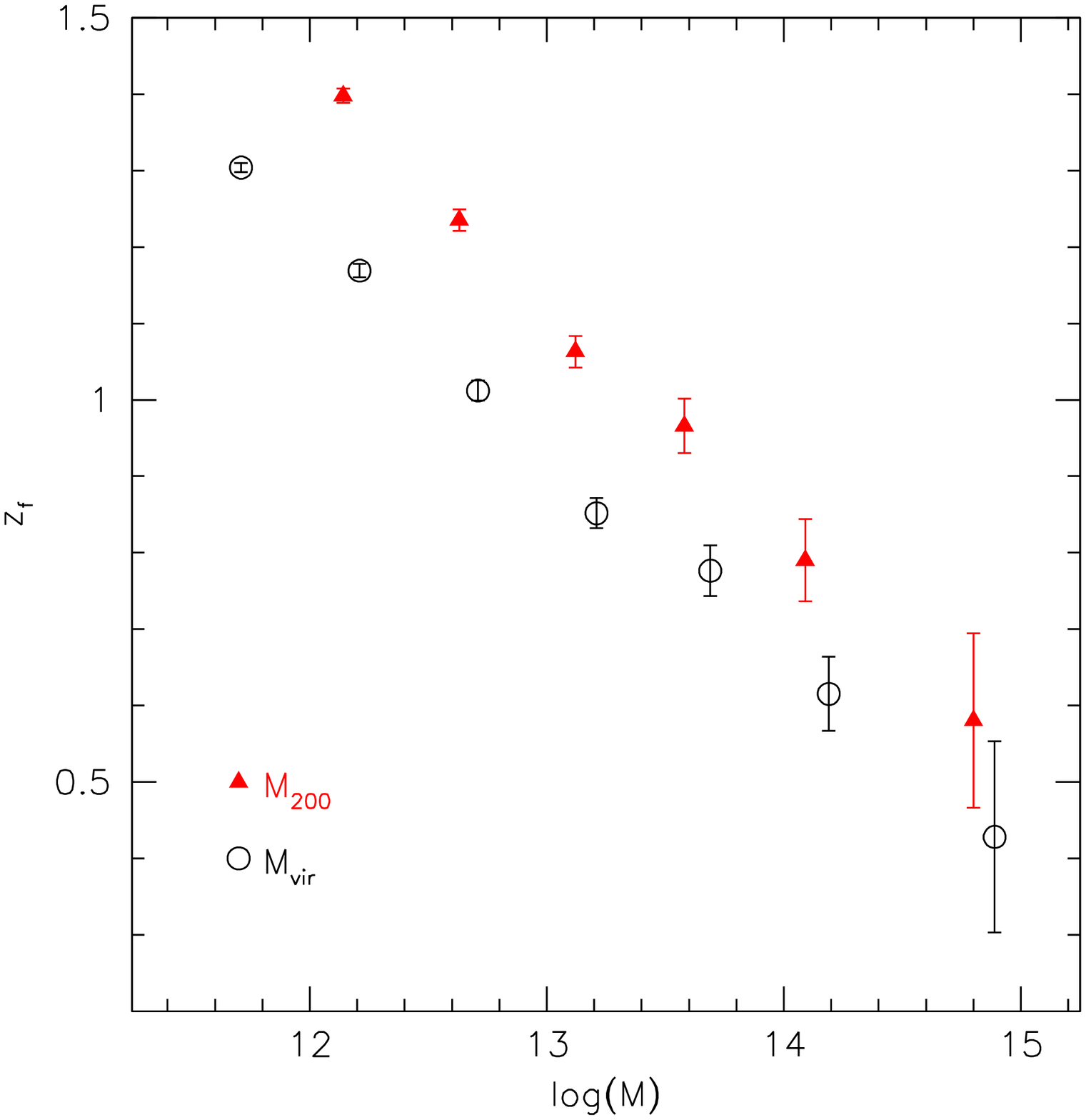}
\caption{Dependence  of formation  time  on definition  of halo  mass:
  haloes defined  to be $200\,\rho_{\rm  crit}(z)$ will have  formed at
  higher  redshifts with  respect  to objects  which are  $\Delta_{\rm
    vir}(z)<200$   times  the  critical   density.   Note,   here  $M$
  corresponds to $M_{200}$  in the case of the  solid triangles and to
  $M_{\rm vir}$ in the case of the open circles. \label{zfvirzf200}}
\end{figure}

In addition, we  argued in the main text  that because $M_{200}<M_{\rm
vir}$, the host  halo formation times for haloes  of a given $M_{200}$
should be  higher than those for  haloes that have  the same numerical
value for  $M_{\rm vir}$.  Figure~\ref{zfvirzf200} shows  that this is
indeed the case.   As a result, we generically  expect $M_{200}$ haloes
at the  present time  to have less  substructure than haloes  with the
same  numerical value  of  $M_{\rm vir}$  (for  comparison see  Figure
\ref{subuv}).

\section{Rescaling Mass Definitions}

In the spherical collapse model,  the density within the virial radius
$R_{\rm  vir}$  is  assumed  to  be $\Delta_{\rm  vir}(z)$  times  the
critical  density at  the  time of  virialization, where  $\Delta_{\rm
  vir}$ is given by equation~(\ref{felix}).  In principle, this can be
used as a  criterion to identify bound objects.   However, because the
details involved  in approximating the timescale  to virialization are
not well understood,  this density is not the only  one which has been
used to identify haloes in simulations.  For example \citet{diemand07}
and related  papers, and \citet{dolag},  define haloes as  being $200$
times  the   mean  background  density.    The  corresponding  radius,
$R_{200,b}$,  is larger  than $R_{\rm  vir}$ at  $z=0$.  On  the other
hand, \citet{swtk01,getal04,speal05} and  related papers define haloes
as  having average densities  $200$ times  the critical  density.  The
corresponding  radius $R_{200,c}$  is  smaller than  $R_{\rm vir}$  at
$z=0$.   Note that  all three  definitions become  comparable  at high
redshift.

For each definition, the enclosed mass can be read as:
\begin{equation}
  M_i = \frac{4 \pi}{3} R_i^3 \Delta_i(z) \rho_c\,, \label{eqrvmv}
\end{equation}
where  $i=\left\{(vir),\,(200,b),\,(200,c)  \right\}$.
\label{massris}
\begin{figure}
\centering
\includegraphics[width=\hsize]{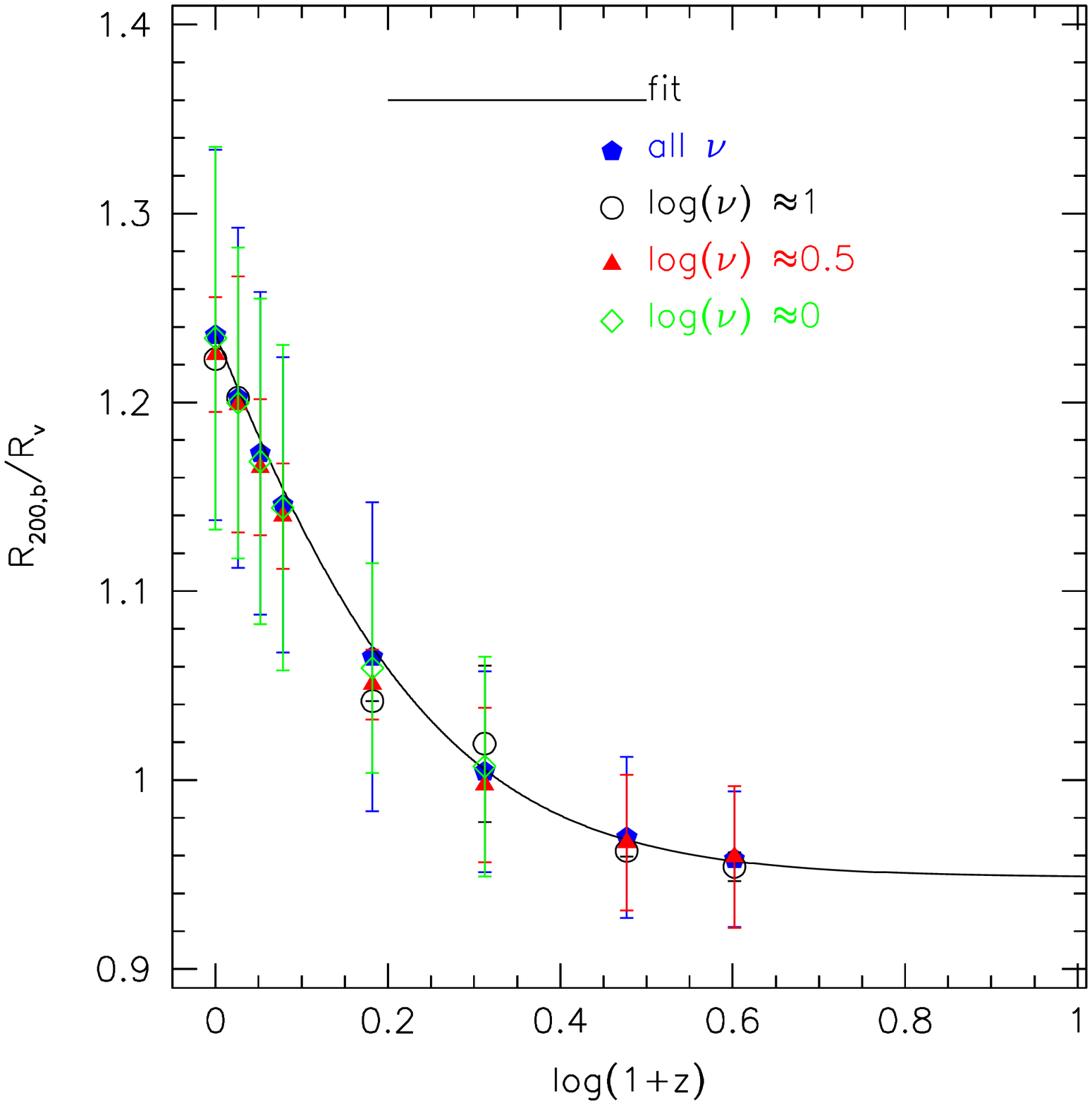}
\caption{\label{ap1}Evolution   of   the  ratio   $R_{200,b}/R_{\rm vir}$.
  Symbols  show  the  simulation   data  for  different  masses  (here
  expressed as $\nu$); solid curve shows equation~\eqref{eqr200b}.}
\end{figure}

\begin{figure}
\centering
\includegraphics[width=\hsize]{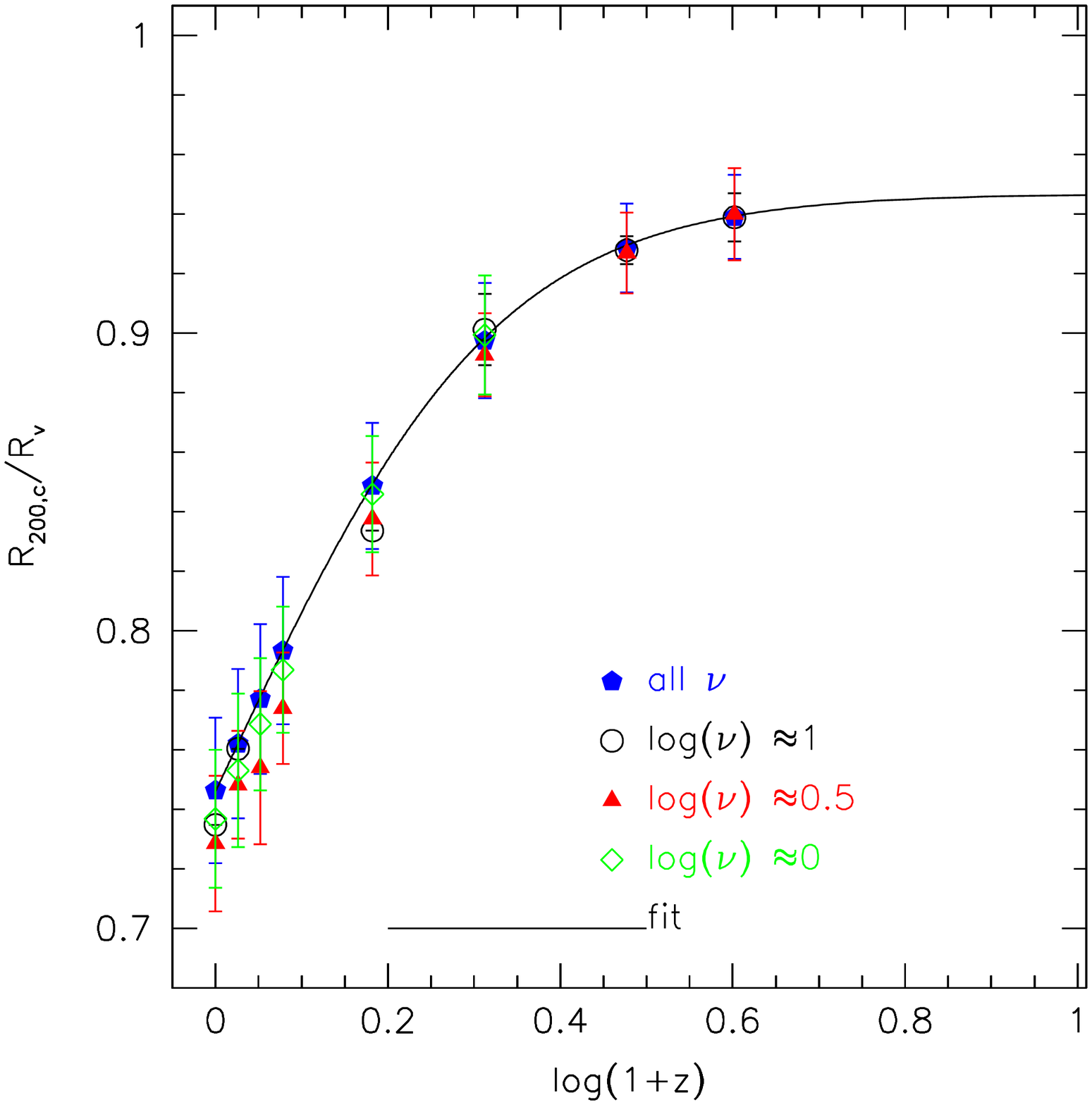}
\caption{Evolution of the ratio $R_{200,c}/R_{\rm vir}$.  Symbols are as 
         in Figure \ref{ap1}; solid curve shows equation~\eqref{eqr200c}. 
         \label{ap2}}
\end{figure}

To    rescale   the   different    mass   definitions    we   estimate
$R_{200,b}/R_{\rm vir}$ and $R_{200,c}/R_{\rm  vir}$ at various $z$ by
computing $R_{200,c}$ and $R_{200,b}$  from the density profile of our
selected haloes (which were defined  to have sizes $R_{\rm vir}$).  We
consider  $8$  simulation outputs  between  $z=0$  and $z=3$.   Figure
\ref{ap1} shows  the evolution  of the ratio  $R_{200,b}/R_{\rm vir}$.
The filled pentagons  show the mean value of  this ratio when averaged
over haloes  of all  masses; the other  data points are  for different
mass bins,  here expressed as  $\nu = \delta_c^2(z)/S(M)$.   The error
bars show  the rms scatter  of the distribution  at each $z$  for each
halo sample.  The solid curve corresponds to:
\begin{equation}
  \frac{R_{200,b}}{R_{\rm vir}} = 1.236 
  \left[ \frac{\Delta_{\rm vir}(z)}{\Delta_{\rm vir}(0)} \right]^{-0.438}\,. 
    \label{eqr200b}
\end{equation}
where $\Delta_{\rm vir}(z)$ is given by equation~\eqref{felix} in the main
text.  The rms  scatter between this curve and  the measured points is
$\sigma_{rms} = 1.1\%$.

Figure \ref{ap2}  shows the  evolution of the  ratio $R_{200,c}/R_{\rm
vir}$.  Data points and error  bars are like in Figure \ref{ap1}.  The
solid curve corresponds to:
\begin{equation}
  \frac{R_{200,c}}{R_{\rm vir}} = 0.746 
  \left[ \frac{\Delta_{\rm vir}(z)}{\Delta_{\rm vir}(0)} \right]^{0.395}\,, 
    \label{eqr200c}
\end{equation}
that well fit  the mean value computed considering  all haloes, with a
$\sigma_{rms} = 0.08\%$.

For consistency we check that the ratio $R_{200,b}/R_{200,c}$ approach
to unity at  high redshift where the universe  is matter dominated and
the dark matter density approaches the critical one.

Equations~\eqref{eqr200b}  and~\eqref{eqr200c}  are  also  useful  for
rescaling different  definitions of halo  concentration.  Halo density
profiles are  will fit by  the functional form of  \citet{nfw97}; this
form  has a  characteristic  scale $r_s$,  where  the density  profile
changes slope, and  the concentration is defined as  the ratio of this
scale to  the virial radius.  Thus,  depending on the  density used to
define   the  halo,   one   might  have   $c_{200}  =  R_{200,c}/r_{\rm s}$
\citep{nfw97,neto,gao08},  $c_{200,b}=R_{200,b}/r_{\rm s}$  \citep{dolag} or
$c_{\rm vir}=R_{\rm vir}/r_{\rm s}$ \citep{bu01}.  Because $r_s$ itself is a
physical quantity  that does not  depend on the halo  mass definition,
the expressions above  can be used to transform  between the different
definitions of concentration.

\label{lastpage}
\end{document}